\documentclass[prd,showpacs,amssymb,nofootinbib]{revtex4}
\usepackage{axodraw,graphicx}
\begin{document}
\topmargin0.0in
\title{Neutrino Propagation in a Strongly Magnetized Medium}
\author{E. Elizalde}
\affiliation{Institute for Space Studies of Catalonia, IEEC/CSIC,\\
 Edifici Nexus, Gran Capit\`a 2-4, 08034 Barcelona, Spain}
\altaffiliation{Presently on leave at: Department of Mathematics,
Massachusetts Institute of Technology, 77 Massachusetts Avenue,
Cambridge, MA 02139-4307}
\author{Efrain J. Ferrer}
\author{Vivian de la Incera}
\affiliation{Physics Department, State University of New York at
Fredonia, Fredonia, NY 14063, USA}

\begin{abstract}
We derive general expressions at the one-loop level for the
coefficients of the covariant structure of the neutrino
self-energy in the presence of a constant magnetic field. The
neutrino energy spectrum and index of refraction are obtained for
neutral and charged media in the strong-field limit ($M_{W}\gg
\sqrt{B}\gg m_{e},T,\mu ,\left| \mathbf{p}\right| $) using the
lowest Landau level approximation. The results found within the
lowest Landau level approximation are numerically validated,
summing in all Landau levels, for strong $B\gg T^{2}$ and
weakly-strong $B \gtrsim T^{2}$ fields. The neutrino energy in
leading order of the Fermi coupling constant is expressed as the
sum of three terms: a kinetic-energy term, a term of interaction
between the magnetic field and an induced neutrino magnetic
moment, and a rest-energy term. The leading radiative correction
to the kinetic-energy term depends linearly on the magnetic field
strength and is independent of the chemical potential. The other
two terms are only present in a charged medium. For strong and
weakly-strong fields, it is found that the field-dependent
correction to the neutrino energy in a neutral medium is much
larger than the thermal one. Possible applications to cosmology
and astrophysics are considered. \pacs{13.15.+g, 14.60.Pq,
95.30.Cq, 98.80.Cq}
\end{abstract}

\maketitle

\section{Introduction}

There are many astrophysical systems on which the physics of
neutrinos in a magnetic field plays an important role. Let us
recall that proto-neutron
stars typically possess very strong magnetic fields. Large magnetic fields $%
B=10^{12}-10^{14}$  G have been associated with the surface of
supernovas \cite{Ginzburg} and neutron stars \cite{Fushiki}, and fields perhaps as large as $%
10^{16}$  G with magnetars \cite{Duncan}. Even larger fields could
exist in the star's interior. It is presumed from the scalar
virial theorem \cite{Lai} that the interior field in neutron stars
could be as high as $10^{18}$  G. A magnetic field as this ($\sim
10^{18}$  G) in the interior of a compact star will be larger in
two orders than the chemical potential characterizing its quark
matter density.

Unveiling the interconnection between the star magnetic field and
its particle current flows could shed new light to the question of
the star evolution. For example, it is well known that neutrinos
drive supernova dynamics from beginning to end. Neutrino emission
and interactions play a crucial role in core collapse supernovae
\cite{SuperN}. Their eventual emission from the proto-neutron star
contains nearly all the energy released in the star explosion.
Neutrino luminosity, emissivity and the specific heat of the
densest parts of the star are governed by charged and neutral
current interactions involving matter at high densities and in the
presence of strong magnetic fields. Thus, a total understanding of
the star cooling mechanism in a strongly magnetized medium is
crucial for astrophysics.

On the other hand, the explanation of large-scale magnetic fields
observed in a number of galaxies, and in clusters of galaxies
\cite{Kronberg} seems to require the existence of seed fields of
primordial origin \cite{Elect-W}.
According to several mechanisms \cite{Grasso}, strong primordial fields $%
\sim 10^{24}$  G could be generated at the electroweak transition.
Even larger fields have been associated with superconducting
magnetic strings, which would generate fields $\sim 10^{30}$  G in
their vicinity if created after inflation \cite{Witten}.

Were primordial magnetic fields present in the early universe,
they would have had non-trivial consequences for particle-physics
cosmology. For instance, as it is well known, oscillations between
neutrino flavors may change the relative abundance of neutrino
species and may thereby affect primordial nucleosynthesis (for a
recent review on neutrinos in cosmology see \cite {Dolgov}).
Therefore, if a strong magnetic field ($M_{W}\gg \sqrt{B}\gg
m_{e},T,\mu ,\left| \mathbf{p}\right| $, with $ M_{W}$ and $m_{e}$
the W-boson and electron masses respectively) modifies the
neutrino energy spectrum of different flavors in different ways, a
primordial magnetic field can consequently influence the
oscillation process in the primeval plasma \cite{Australia}.

The propagation of neutrinos in magnetized media has been
previously investigated by several authors
\cite{McKeon}-\cite{Nieves}. Weak-field calculations were done for
magnetized vacuum in Refs. \cite{McKeon}, \cite{Feldman}, and at $
T\neq 0$ and $\mu =0$ ($\mu $ is the electric chemical potential)
in Refs. \cite{Weak-F}. In the $\mu =0$ case, as long as $B<
T^{2}$, both the field- and temperature-dependent leading
contributions to the neutrino energy resulted of
$1/M_{W}^{4}$-order \cite{McKeon}-\cite{Weak-F}. In the charged
plasma \cite{Nieves}, the field-dependent terms were much larger,
$\sim 1/M_{W}^{2}$, but they vanished in the ($\mu \rightarrow
0$)-limit. The weak-field results of papers
\cite{McKeon}-\cite{Nieves} led to think that the magnetic-field
effects could be significant in astrophysics, because of the
field- and $\mu$-dependent terms of order $1/M_{W}^{2}$, but were
irrelevant ($\sim 1/M_{W}^{4}$-order) in the early universe due to
its charge-symmetric character ($\mu=0$). Hence, it has been
assumed that in cosmology the main contribution to neutrino energy
was the purely thermal term of order $T^{4}/M_{W}^{4}$
\cite{Raffelt}. However, as we will show below, a strong magnetic
field ($M_{W}\gg \sqrt{B}\gg m_{e},T,\mu ,\left| \mathbf{p}\right|
$) gives rise to a new contribution to the neutrino energy that is
linear in the field, independent of the chemical potential, and
that is of the same order ($1/M_{W}^{2}$) as the largest terms
found in the weak-field charged-medium case. This new result can
turn magnetic-field effects relevant for cosmology.

In recent papers \cite{Elizalde}, \cite{Efrain}, we investigated
the effects of a strong magnetic field on neutrinos in magnetized
vacuum (i.e. with $T=0$ and $\mu =0$). There, to facilitate the
calculations in the strong-field limit, we extended the Ritus'
Ep-eigenfunction method of diagonalization of the Green functions
of spin-1/2 charged particle in electromagnetic field
\cite{Ritus}-\cite{Ritus-Book}, to the case of spin-1 charged
particles \cite {Elizalde}. This formulation, which is
particularly advantageous for strong-field calculations, provides
an alternative method to the Schwinger approach to address QFT
problems in electromagnetic backgrounds \cite {Schwinger}. The use
of the Ritus' method resulted very convenient to study the
neutrino-self-energy in magnetized media, since it allowed to
diagonalize in momentum space both the electron and the W-boson
Green's functions in the presence of a magnetic field. Ritus'
formalism has also been recently applied to investigate
non-perturbative QFT in electromagnetic backgrounds \cite{Ng}.

From the above considerations it is clear that strong magnetic
fields can play a significant role in a variety of astrophysical
systems, and possibly also in the early universe. For these
applications, the analysis has to be carried out in the presence
of a medium. Thus, in the present paper we extend the results
obtained in papers \cite{Elizalde} and \cite{Efrain} to include
finite temperature and density, performing a detailed study of the
effects of a strong magnetic field on neutrino propagation in
neutral and charged media, and discussing possible applications to
astrophysics and cosmology.

We stress that in calculating the neutrino self-energy in a
magnetized medium, we should consider, as usual, its vacuum and
statistical parts. In this case the vacuum part depends on the
magnetic field, and for strong fields it can make important
contributions even at high temperatures $T^{2}\gtrsim eB$. The
reason is that the vacuum and statistical terms have different
analytical behaviors, due to the lack of the statistic ultraviolet
cutoff in the vacuum part. This fact gives rise to a
field-dependent vacuum contribution ($1/M_{W}^{2}$-order) which is
larger than the thermal one ($1/M_{W}^{4}$-order), into the
self-energy. Therefore, as shown in this paper, a strong magnetic
field can become more relevant than temperature for neutrino
propagation in neutral media.

The plan of the paper is as follows. In
Sec.~\ref{selfen-gen-struc} we consider the radiative correction
to the neutrino dispersion relation in the presence of a constant
magnetic field. We introduce the general covariant structure of
the neutrino self-energy in the presence of an external field, and
find the dispersion relation as a function of the coefficients of
each independent term in the covariant structure. The general form
of the found dispersion relation goes beyond any given
approximation and medium characteristic and serves as a guidance
for particular applications. The general expressions for the
coefficients of the self-energy structures in the presence of the
magnetic field are found in the one-loop approximation in
Sec.~\ref{1-loop-selfen}. The leading behavior in the
$1/M_{W}^{2}$ expansion of those coefficients are then calculated
in Sec.~\ref{strong-b-selfen} in the strong-field limit (i.e. in
the lowest Landau level (LLL) approximation) for neutral and
charged magnetized media. These results are then used in
Sec.~\ref{index-ref} to find the corresponding neutrino dispersion
relations and the indexes of refraction in neutral and charged
strongly magnetized media. In Sec.~\ref{numerical}, the
LLL-approximation is numerically corroborated by summing in all
Landau levels and finding the values of the coefficients for
parameter ranges corresponding to strong $B \gg T^{2}$ and weakly
strong $B \gtrsim T^{2}$ fields. Possible applications to
cosmology and astrophysics are discussed in Sec.~\ref{applicat}.
Finally, in Sec.~\ref{conclusions} we summarize the main outcomes
of the paper and make some final remarks.

\section{Neutrino Self-Energy General Structure in a Magnetic
Field}\label{selfen-gen-struc}

The neutrino field equation of motion in a magnetized medium,
including radiative corrections, is
\begin{equation}
\left[ p\llap/+\sum \right]\Psi _{L}=0  \label{1}
\end{equation}
where the neutrino self-energy operator $\sum (p)$ depends on the
parameters characterizing the medium, as for instance,
temperature, magnetic field, particle density, etc.

The operator $\sum (p)$ is a Lorentz scalar that can be formed in
the spinorial space taking the contractions between the
characteristic vectors and tensors of the system with all the
independent elements of the Dirac ring. Explicit chirality reduces
it to
\begin{equation}
\sum =R\overline{\sum }L,\qquad \overline{\sum }=V_{\mu }\gamma
^{\mu } \label{2}
\end{equation}
where $L,R=\frac{1}{2}(1\pm \gamma _{5})$ are the chiral projector
operators, and $V_{\mu }$ is a Lorentz vector that can be spanned
as a superposition of four basic vectors that can be formed from
the characteristic tensors of the problem. In a magnetized medium,
besides the neutrino four-momentum $p_{\mu \text{ }}$, we have to
consider the magnetic-field strength tensor $F_{\mu \nu }$, to
form the covariant structures
\begin{equation}
\overline{\sum }(p,B)=ap\llap/_{\Vert }+bp\llap/_{\perp }+cp^{\mu }\widehat{%
\widetilde{F}}_{\mu \nu }\gamma ^{\nu }+idp^{\mu }\widehat{F}_{\mu
\nu }\gamma ^{\nu }.  \label{3}
\end{equation}
In (\ref{3}) we introduced the notations $\widehat{F}_{\mu \nu
}=\frac{1}{ \left| B\right| }F_{\mu \nu }$, and
$\widehat{\widetilde{F}}_{\mu \nu }=$ $ \frac{1}{2\left| B\right|
}\varepsilon _{\mu \nu \rho \lambda }F^{\rho \lambda }$. In the
covariant representation the magnetic field can be expressed as
$B_{\mu }=\frac{1}{2}\varepsilon _{\mu \nu \rho \lambda }u^{\nu
}F^{\rho \lambda }$, where $u_{\mu }$ is the vector four-velocity
of the center of mass of the magnetized medium. The presence of
the magnetic field, and hence of the dimensionless magnetic field
tensor $\widehat{F}_{\mu \nu }$ and its dual
$\widehat{\widetilde{F}}_{\mu \nu }$, allows the covariant
separation in (\ref{3}) between longitudinal and transverse
momentum terms that appears naturally in magnetic backgrounds
\begin{equation}
p\llap/_{\Vert }=p^{\mu }\widehat{\widetilde{F}}_{\mu}\,^{\rho }
\widehat{\widetilde{F}}_{\rho\nu }\gamma ^{\nu },\qquad
p\llap/_{\perp }=p^{\mu } \widehat{F}_{\mu }\,^{\rho
}\widehat{F}_{\rho \nu }\gamma ^{\nu }. \label{4}
\end{equation}
Coefficients $a$, $b$, $c$, and $d$ are Lorentz scalars that
depend on the parameters of the theory and the approximation used.
Notice that finite temperature and/or density would not introduce
any new term into Eq. (\ref{3}). This formulation is indeed the
most general one for a magnetized medium. In magnetized space,
even if $T=\mu =0$, the vector $u_{\mu }$ is present, since the
presence of a constant magnetic field fixes a special Lorentz
frame: the rest frame (on which $u_{\mu }=(1,0,0,0)$) where the
magnetic field is defined ($u_{\mu }F^{\mu \nu }=0)$. Thus, the
inclusion of a medium does not break any additional symmetry. To
express $\overline{\sum }(p,B)$ in terms of $u_{\mu }$ only means,
therefore, to make a change in the selected basis
vectors\footnote{This point will turn clear in Sec.
\ref{strong-b-selfen} when expressing $\overline{\sum }(p,B)$ in
terms of $u_{\mu }$ in the charged medium.}.

The neutrino energy spectrum in a magnetic background can be found
from Eq. (\ref{1}) as the nontrivial solution of
\begin{equation}
\det \left[ p\llap/+\sum (p,B)\right] =0.  \label{5}
\end{equation}

Using the covariant structure (\ref{3}), the dispersion relation
(\ref{5}) takes the form
\begin{equation}
p_{0}^{2}=p_{3}^{2}+\frac{ \left( 1+b\right) ^{2}-d^{2} }{\left(
1+a\right) ^{2}-c^{2} } \, p_{\bot }^{2}. \label{6}
\end{equation}

One of the main goals of this paper is to find the coefficients
$a$, $b$, $c$, and $d$, in the one-loop approximation, and hence,
the dispersion relations for different systems in the presence of
strong magnetic-field backgrounds.

\section{One-Loop Neutrino Self-Energy}\label{1-loop-selfen}

Let us consider the one-loop corrections to the neutrino
self-energy in the presence of a constant magnetic field. To
leading order in the Fermi coupling constant the main
field-dependent contribution to the self-energy comes from the
bubble diagram (Fig.I-a) with internal lines of virtual charged
leptons and W-bosons, and from the tadpole diagram (Fig.I-b) with
virtual loop of charged leptons.

In a neutral medium only the bubble diagram contributes, while in
a charged medium both diagrams should be considered. However, only
the bubble distinguishes between neutrino flavors, since the
flavor of the internal charged lepton is directly associated to
the flavor of the propagating neutrino (i.e. the one appearing in
the external legs). For the tadpole, because the internal boson is
the neutral Z-boson, the charged leptons corresponding to
different families are linked to the same neutrino flavor.
Therefore, in regards to magnetic field effects on neutrino
oscillations, the bubble contribution is the essential one. Thus,
because the ultimate goal of our study is to elucidate the
possible effect of a strong magnetic field on neutrino flavor
oscillations, henceforth we only consider the bubble diagram
contribution, represented in Fig.I-a, into the neutrino
self-energy.

The bubble contribution to the one-loop self-energy is
\begin{equation}
\Sigma (x,y)=\frac{ig^{2}}{2}R\gamma _{\nu }S(x,y)\gamma ^{\mu
}G_{F}(x,y)_{\mu }\,^{\nu }L  \label{7}
\end{equation}
where $S(x,y)$ and $G_{F}(x,y)_{\nu }\,^{\mu }$ are the Green's
functions of the electron and W-boson, in the presence of the
magnetic field, respectively. Since both virtual particles are
electrically charged, the magnetic field interacts with both of
them producing the Landau quantization of the corresponding
transverse momenta \cite{Elizalde}, \cite{Efrain}.

\begin{center}\begin{picture}(450,80)(0,-10)
\Vertex(180,10){1.5} \Vertex(120,10){1.5} \SetColor{Black}
\Line(120,10)(180,10)\SetColor{Red}\DashLine(100,10)(120,10){5}
\SetColor{Red}\DashLine(180,10)(200,10){5}\GCirc(300,50){13}{1}
\SetColor{Blue}\ZigZag(300,37)(300,10){3}{5}\Vertex(300,37){1.5}
\Vertex(300,10){1.5} \SetColor{Red}\DashLine(280,10)(320,10){5}
\Text(150,0)[]{($a$)}\Text(300,0)[]{($b$)}\SetColor{Green}
\PhotonArc(150,10)(30,0,180){4}{8.5}
\end{picture}\end{center}

Fig.I. One-loop contributions to the neutrino self-energy in a
magnetized medium. (a) Bubble graph - The dashed line represents
the test neutrino, the solid line a charged lepton of the same
family of the test neutrino, and the wiggly line the W-boson. (b)
Tadpole graph - The dashed line represents the test neutrino, the
solid line a charged lepton of any species, and the zigzag line
the Z-boson.
\\

The electron Green's function, diagonal in momentum space at
arbitrary field strength, was obtained in Refs. \cite{Ritus},
\cite{Ritus-Book} by Ritus, using what has become an alternative
method to the Schwinger approach \cite{Schwinger} to deal with QFT
problems on electromagnetic backgrounds. In this method the
electron Green's function in configuration space is given by
\begin{equation}
S(x,y)=\sum_{\it l}\hspace{-0.57cm}\int
\frac{d^{3}\widehat{q}}{\left( 2\pi \right) ^{4}}E_{q}\left(
x\right) \frac{1}{\gamma .\overline{q}+m_{e}}\overline{E}_{q}(y)
\label{8}
\end{equation}
where $\widehat{q}_{\mu }=(q_{0},0,q_{2},q_{3})$,
$\overline{E}_{q}\equiv\gamma^{0}E^{\dag}_{q}\gamma^{0}$, and the
magnetic field has been specialized in the rest frame along the
$\mathcal{Z}$-direction (i.e. given in the Landau gauge as $A_{\mu
}^{ext}=Bx_{1}\delta _{\mu 2}$).

The transformation functions $E_{q}(x)$ in (\ref{8}) play the
role, in the presence of magnetic fields, of the usual Fourier
functions $e^{iqx}$ in the free case and are given by
\begin{equation}
E_{q}(x)=\sum\limits_{\sigma }E_{q\sigma }(x)\Delta (\sigma ),
\label{9}
\end{equation}
with spin matrix
\begin{equation}
\Delta (\sigma )=diag(\delta _{\sigma 1},\delta _{\sigma
-1},\delta _{\sigma 1},\delta _{\sigma -1}),\qquad \sigma =\pm 1,
\label{10}
\end{equation}
and eigenfunctions
\begin{equation}
E_{q\sigma }(x)=N(\kappa)\exp i(\widehat{q}\cdot
\widehat{x})D_{\kappa}(\rho ) \label{11}
\end{equation}
where $N(\kappa)=(4\pi \left| eB\right|
)^{\frac{1}{4}}/\sqrt{\kappa!}$ is a normalization factor and
$D_{\kappa}(\rho )$ denotes the parabolic cylinder functions
\cite{handbook} with argument $\rho =\sqrt{2\left| eB\right|
}(x_{1}-\frac{q_{2}}{eB})$ and positive integer index
\begin{equation}
\kappa=\kappa(\mathit{l},\sigma )\equiv \mathit{l} +\frac{\sigma
}{2}-\frac{1}{2} ,\;\quad \kappa=0,1,2,...  \label{12}
\end{equation}
The integer $\mathit{l} $ in Eq. (\ref{12}) labels the Landau
levels $\mathit{l}=0,1,2,...$.

The electron momentum eigenvalue in the presence of the magnetic
field, $\overline{q}$, is given by
\begin{equation}
\overline{q}_{\mu }=\left( q_{0},0,-sgn\left( eB\right)
\sqrt{2\left| eB\right| \mathit{l}},q_{3}\right)  \label{13}
\end{equation}

Ritus' technique for spin-1/2 particles was recently extended to
the spin-1 particle case in Refs.~\cite{Elizalde}, \cite{Efrain}.
In this case the corresponding W-boson Green's function can be
diagonalized in momentum space as
\begin{equation}
G_{F}(k,k^{\prime })_{\mu }\,^{\nu }=\left( 2\pi \right)
^{4}\widehat{\delta }^{(4)}(k-k^{\prime })\frac{\delta _{\mu
}\,^{\nu }}{\overline{k} ^{2}+M_{W}^{2}}  \label{14}
\end{equation}
where $\widehat{\delta }^{(4)}(k-k^{\prime
})={\delta}^3(k-k^{\prime }){\delta}_{nn'}$, and the W-boson
momentum eigenvalue in the presence of the magnetic field,
$\overline{k}$, is given by the relation

\begin{equation}
\overline{k}^{2}=-k_{0}^{2}+k_{3}^{2}+2(\mathit{n}-1/2)eB,\qquad \mathit{n}%
=0,1,2,...  \label{15}
\end{equation}
with $\mathit{n}$ denoting the W-boson Landau levels. Considering the mass-shell condition $\overline{k}^{2}=-M_{W}^{2}$ in Eq. (%
\ref{15}), we can identify at $eB>M_{W}^{2}$ the so called
``zero-mode problem" \cite{Savvidy}-\cite{Olesen}. As known, at
those values of the magnetic field a vacuum instability appears
giving rise to W-condensation
\cite{Olesen}. In our calculations we restrict the magnitude of the magnetic field to $%
eB<M_{W}^{2} $, so to avoid the presence of any tachyonic mode.

The W-boson Green's function in configuration space is then
\begin{equation}
G_{F}(x,y)_{\mu }\,^{\nu }= \sum_{\it n}\hspace{-0.57cm}\int
\frac{d^{3}\widehat{k}}{\left( 2\pi \right) ^{4}}\Gamma
_{k}^{\alpha }\,_{\mu }\left( x\right) \frac{\delta _{\alpha
}\,^{\beta }}{\overline{k} ^{2}+M_{W}^{2}}\Gamma _{\,k\beta
}^{\dagger }\,^{\nu }(y) \label{16}
\end{equation}
where the transformation functions are
\begin{equation}
\Gamma _{k}^{\alpha }\,_{\mu }\left( x\right) =P^{\alpha
}\,_{\gamma }\left[ \mathcal{F}_{k}\left( x\right) \right]
^{\gamma }\,_{\lambda } P^{-1\lambda }\,_{\mu } \label{17}
\end{equation}
with
\begin{equation}
P^{\alpha }\,_{\gamma }=\frac{1}{\sqrt{2}}\left(
\begin{array}{llll}
\sqrt{2} & 0 & 0 & 0 \\
0 & 1 & 1 & 0 \\
0 & i & -i & 0 \\
0 & 0 & 0 & \sqrt{2}
\end{array}
\right)  \label{18}
\end{equation}
and
\begin{equation}
\left[ \mathcal{F}_{k}\left( x\right) \right] ^{\gamma
}\,_{\lambda }=\sum\limits_{\eta =0,\pm 1}\mathcal{F}_{k\eta
}(x)\left[ \Omega ^{(\eta )}\right] ^{\gamma }\,_{\lambda }
\label{19}
\end{equation}
The matrices $\Omega ^{(\eta )}$ are explicitly given in terms of
the W-boson spin projections $\eta$ by
\begin{equation}
\Omega ^{(\eta )}=diag(\delta _{\eta ,0},\delta _{\eta ,1},\delta
_{\eta ,-1},\delta _{\eta ,0}),\qquad \eta =0,\pm 1  \label{20}
\end{equation}
and the eigenfunctions $\mathcal{F}_{k\eta }$ are
\begin{equation}
\mathcal{F}_{k\eta }(x)=N(\kappa')\exp i(\widehat{k}\cdot
\widehat{x})D_{\kappa'}(\xi ) \label{21}
\end{equation}
with $N(\kappa')$ a normalization factor similar to that appearing
in (\ref{11}), and $D_{\kappa'}(\xi )$ denoting the parabolic
cylinder functions with argument $\xi =\sqrt{2\left| eB\right|
}\left( x_{1}-k_{2}/eB\right)$ and positive integer index
$\kappa'$ given in terms of the W-boson Landau numbers
$\mathit{n}$ and spin projections $\eta$ as

\begin{equation}
\kappa'=\kappa'(\mathit{n},\eta )\equiv \mathit{n} -\eta -1
,\;\quad \kappa'=0,1,2,...  \label{21-a}
\end{equation}

As the neutrino is an electrically neutral particle, the
transformation to momentum space of its self-energy can be carried
out by the usual Fourier transform
\begin{equation}
\left( 2\pi \right) ^{4}\delta ^{(4)}(p-p^{\prime })\Sigma
(p,B)=\int d^{4}xd^{4}ye^{-i(p.x-p^{\prime }.y)}\Sigma (x,y)
\label{22}
\end{equation}

Substituting with (\ref{7}), (\ref{8}) and (\ref{16}) in the RHS
of (\ref{22}) we obtain

\[
\left( 2\pi \right) ^{4}\delta ^{(4)}(p-p^{\prime })\Sigma (p,B)=\frac{%
ig^{2}}{2}\int d^{4}x\int d^{4}ye^{-i(p.x-p^{\prime }.y)}\left\{
R\left[ \gamma _{\nu }\left(
\sum_{\it l}\hspace{-0.57cm}\int%
\frac{d^{3}\widehat{q}}{\left( 2\pi \right) ^{4}}E_{q}\left( x\right) \frac{1%
}{\gamma .\overline{q}+m_{e}}\overline{E}_{q}(y)\right) \right.
\right.
\]
\begin{equation}
\left. \left. \gamma ^{\mu }\left(
\sum_{\it n}\hspace{-0.5cm}\int%
\frac{d^{3}\widehat{k}}{\left( 2\pi \right) ^{4}}\frac{\Gamma
_{k}^{\alpha
}\,_{\mu }\left( x\right) \Gamma _{\,k\alpha }^{\dagger }\,^{\nu }(y)}{%
\overline{k}^{2}+M_{W}^{2}}\right) \right] L\right\}  \label{23}
\end{equation}

The coefficients of the covariant expression for $\Sigma (p,B)$
can be found using the explicit form (\ref{23}) in the covariant
expression
\begin{equation}
\left( 2\pi \right) ^{4}\delta ^{(4)}(p-p^{\prime })\Sigma
(p,B)=a^{\prime }p\llap/_{\Vert }+b^{\prime }p\llap/_{\perp
}+c^{\prime }p^{\mu }\widehat{\widetilde{F}}_{\mu \nu }\gamma
^{\nu }+id^{\prime }p^{\mu }\widehat{F}_{\mu \nu }\gamma ^{\nu }
\label{24-a}
\end{equation}

Introducing in (\ref{23}) the transformation functions $E_{q}$ and
$\Gamma _{k}^{\alpha }\,_{\mu }$, and taking into account the
following properties of the spinor matrices
\[
\Delta \left( \pm \right) ^{\dagger }=\Delta \left( \pm \right)
,\qquad \Delta \left( \pm \right) \Delta \left( \pm \right)
=\Delta \left( \pm \right) ,\qquad \Delta \left( \pm \right)
\Delta \left( \mp \right) =0
\]
\[
\gamma ^{\shortparallel }\Delta \left( \pm \right) =\Delta \left(
\pm \right) \gamma ^{\shortparallel },\quad \gamma ^{\bot }\Delta
\left( \pm \right) =\Delta \left( \mp \right) \gamma ^{\bot },
\]
\begin{equation}
\quad L\Delta \left( \pm \right) =\Delta \left( \pm \right)
L,\quad R\Delta \left( \pm \right) =\Delta \left( \pm \right) R
\label{24}
\end{equation}
where the notation $\gamma ^{\shortparallel }=(\gamma ^{0},\gamma
^{3})$ and $\gamma ^{\bot }=(\gamma ^{1},\gamma ^{2})$ was used,
we find for the coefficients appearing in (\ref{24-a}) the
following general expressions in the one-loop approximation
(Henceforth we consider $sgn(eB)>0$, to simplify the notation.)
\[
a^{\prime }=\frac{ig^{2}}{2p_{\Vert }^{2}}\sum\limits_{l,n}\int
d^{4}x\int d^{4}y\int \frac{d^{3}\widehat{q}}{\left( 2\pi \right)
^{4}}\int \frac{d^{3}\widehat{k}}{\left( 2\pi \right)
^{4}}\frac{e^{-i(p.x-p^{\prime }.y)}}{\left(
\overline{q}^{2}+m_{e}^{2}\right) \left( \overline{k}
^{2}+M_{W}^{2}\right) }
\]
\begin{equation}
\left\{
(q_{0}+q_{3})(p_{3}-p_{0})I_{n-2,l-1}(x)I_{n-2,l-1}^{*}(y)+(q_{3}-q_{0})(p_{3}+p_{0})I_{n,l}(x)I_{n,l}^{*}(y)\right\}
\label{25}
\end{equation}
\[
b^{\prime }=\frac{ig^{2}}{2p_{\bot }^{2}}\sum\limits_{l,n}\int
d^{4}x\int d^{4}y\int \frac{d^{3}\widehat{q}}{\left( 2\pi \right)
^{4}}\int \frac{d^{3}\widehat{k}}{\left( 2\pi \right)
^{4}}\frac{e^{-i(p.x-p^{\prime }.y)}}{\left(
\overline{q}^{2}+m_{e}^{2}\right) \left( \overline{k}
^{2}+M_{W}^{2}\right) }
\]
\begin{equation}
\left\{ (p_{2}+ip_{1})\overline{q}%
_{2}I_{n-1,l}(x)I_{n-1,l-1}^{*}(y)+(p_{2}-ip_{1})\overline{q}
_{2}I_{n-1,l-1}(x)I_{n-1,l}^{*}(y)\right\}  \label{26}
\end{equation}
\[
c^{\prime }=\frac{ig^{2}}{2p_{\Vert }^{2}}\sum\limits_{l,n}\int
d^{4}x\int d^{4}y\int \frac{d^{3}\widehat{q}}{\left( 2\pi \right)
^{4}}\int \frac{d^{3} \widehat{k}}{\left( 2\pi \right)
^{4}}\frac{e^{-i(p.x-p^{\prime }.y)}}{\left(
\overline{q}^{2}+m_{e}^{2}\right) \left( \overline{k}
^{2}+M_{W}^{2}\right) }
\]
\begin{equation}
\left\{
(q_{0}+q_{3})(p_{3}-p_{0})I_{n-2,l-1}(x)I_{n-2,l-1}^{*}(y)-(q_{3}-q_{0})(p_{3}+p_{0})I_{n,l}(x)I_{n,l}^{*}(y)\right\}
\label{27}
\end{equation}
\[
d^{\prime }=\frac{-ig^{2}}{2p_{\bot }^{2}}\sum\limits_{l,n}\int
d^{4}x\int d^{4}y\int \frac{d^{3}\widehat{q}}{\left( 2\pi \right)
^{4}}\int \frac{d^{3}\widehat{k}}{\left( 2\pi \right)
^{4}}\frac{e^{-i(p.x-p^{\prime
}.y)}}{\left( \overline{q}^{2}+m_{e}^{2}\right) \left( \overline{k}%
^{2}+M_{W}^{2}\right) }
\]
\begin{equation}
\left\{
(p_{2}+ip_{1})\overline{q}_{2}I_{n-1,l}(x)I_{n-1,l-1}^{*}(y)-(p_{2}-ip_{1})\overline{q}_{2}I_{n-1,l-1}(x)I_{n-1,l}^{*}(y)\right\}
\label{28}
\end{equation}
where
\begin{equation}
I_{n,l}(x)=2\pi \sqrt{eB}\exp [i\widehat{x}\cdot
(\widehat{k}+\widehat{q})]\varphi _{n}(\xi /\sqrt{2})\varphi
_{l}(\rho /\sqrt{2}) \label{29}
\end{equation}
and $\varphi _{m}(x)$ are the orthonormalized harmonic oscillator
wave functions, defined in terms of the Hermite polynomials
$H_{m}(x)$ as
\begin{equation}
\varphi _{m}(x/\sqrt{2})=\frac{2^{-m/2}}{[\sqrt{\pi
}m!]^{1/2}}H_{m}(x/\sqrt{2})\exp -(x^{2}/4)  \label{30}
\end{equation}

Expressions (\ref{25})-(\ref{28}), when substituted in
(\ref{24-a}), give the general formula for the one-loop neutrino
self-energy in a constant magnetic field of arbitrary strength.
Note that, in this approach, the W-boson/magnetic-field
interaction is kept in the poles of the self-energy operator
through the effective momentum $\overline{k}^{2}$, as well as in
the harmonic oscillator wave functions $\varphi _{n}(\xi
/\sqrt{2})$.

\section{Neutrino Self-Energy in the Strong-Field Limit}\label{strong-b-selfen}

Since in the strong-field limit ($M_{W}\gg \sqrt{B}\gg m_{e},T,\mu
,\left| \mathbf{p}\right| $) the gap between the electron Landau
levels is larger than the rest of the parameters entering in the
electron energy, it is consistent to use the LLL approximation in
the electron Green's function, while in the W-boson Green's
function, because $M_{W}\gg \sqrt{B},$ we must sum in all W-boson
Landau levels.

The neutrino self-energy in a magnetized vacuum was found within
this approximation in Ref.~\cite{Elizalde}. In what follows, we
extend that result to neutral and charged media, introducing
finite temperature and density effects. However, since the vacuum
contribution is always present in the statistical calculations of
the self-energy, we summarize below the results found in
Ref.~\cite{Elizalde}.

In the strong field approximation ($\mathit{l}=0$) the covariant
structure of the neutrino self-energy reduces to
\begin{equation}
\left( 2\pi \right) ^{4}\delta ^{(4)}(p-p^{\prime
})\overline{\Sigma } _{0}(p,B)=a_{0}^{\prime }p\llap/_{\Vert
}+c_{0}^{\prime }p^{\mu }\widehat{\widetilde{F}}_{\mu \nu }\gamma
^{\nu }  \label{31}
\end{equation}
That is, $b^{\prime }$ and $d^{\prime }$ are zero and
\begin{equation}
a_{0}^{\prime }=\frac{ig^{2}(p_{3}+p_{0})}{2p_{\Vert }^{2}}
\sum\limits_{n}\int d^{4}x\int d^{4}y\int
\frac{d^{3}\widehat{q}}{\left( 2\pi \right) ^{4}}\int
\frac{d^{3}\widehat{k}}{\left( 2\pi \right) ^{4}}
\frac{(q_{3}-q_{0})I_{n,0}(x)I_{n,0}^{*}(y)e^{-i(p.x-p^{\prime
}.y)}}{\left( q_{\Vert }^{2}+m_{e}\right) \left(
\overline{k}^{2}+M_{W}^{2}\right) } \label{32}
\end{equation}
\begin{equation}
c_{0}^{\prime }=-a_{0}^{\prime }  \label{33}
\end{equation}

After summing in $\mathit{n}$ and integrating in $x$, $y$, and
$\widehat{k}$ we obtain
\begin{equation}
a_{0}^{\prime }=-c_{0}^{\prime }=\frac{ig^{2}\pi
(p_{3}+p_{0})}{2p_{\Vert }^{2}}e^{-p_{\bot }^{2}/2eB}\int
d^{2}q_{\Vert }\frac{(q_{3}-q_{0})}{[q_{\Vert
}^{2}+m_{e}][(q_{\Vert }-p_{\Vert })^{2}-eB+M_{W}^{2}]} \label{34}
\end{equation}

Integrating in $q_{\Vert }$ and considering that $M_{W}^{2}\gg
eB,m_{e}^{2},p_{\Vert }^{2}$, we obtain for the neutrino
self-energy in the strongly magnetized vacuum
\begin{equation}
\overline{\Sigma }_{0}(p,B)=a_{0}p\llap/_{\Vert }+c_{0}p^{\mu }\widehat{%
\widetilde{F}}_{\mu \nu }\gamma ^{\nu }  \label{34-a}
\end{equation}
with coefficients given by
\begin{equation}
a_{0}=-c_{0}\simeq \frac{-g^{2}eB}{2(4\pi
)^{2}M_{W}^{2}}e^{-p_{\bot }^{2}/2eB}  \label{35}
\end{equation}

Notice the $1/M_{W}^{2}$-order of the leading contribution in the
LLL approximation. Using the zero-temperature weak-field results
of Ref.~\cite {McKeon} we can identify the scalar coefficients of
the general structure (\ref {3}) for that case as
\begin{equation}
\widetilde{a}_{0}=\widetilde{b}_{0}\simeq 0,\qquad
\widetilde{c}_{0}=-\frac{3}{2}\widetilde{d}_{0}\simeq
\frac{6g^{2}eB}{(4\pi )^{2}M_{W}^{2}} \label{36}
\end{equation}

If we compare the vacuum results at weak-field, Eq. (\ref{36}),
with those at strong field, Eq. (\ref{35}), we can see that in
both cases the non-zero coefficients have field-dependent leading
contributions of $1/M_{W}^{2}$-order. However, as we will show in
Sec.~\ref{index-ref}, they will play very different roles in the
neutrino energy spectrum, since they are associated to different
self-energy structures. While the strong-field results (\ref{35})
produce a magnetic-field dependence in the neutrino energy
spectrum which is linear in the Fermi coupling constant, the
weak-field results (\ref{36}) generate a smaller second-order
contribution.

\subsection{Neutral Medium in Strong Magnetic Field}\label{neutral-medium}

In a neutral medium the self-energy operator can be written as
\begin{equation}
\overline{\Sigma }(p,B)=\overline{\Sigma
}_{0}(p,B)+\overline{\Sigma }_{T}(p,B),  \label{37}
\end{equation}
where $\overline{\Sigma }_{0}$ is the vacuum contribution (given
in the strong-field approximation by Eqs.
(\ref{34-a})-(\ref{35})), and $\overline{\Sigma}_{T}$ is the
statistical part which depends on temperature and magnetic field
(in a charged medium the statistical part can also depend, of
course, on the chemical potential). Our goal now is to find
$\overline{\Sigma }_{T}(p,B)$ in the parameter range $M_{W}\gg
\sqrt{B}\gg m_{e},T,\left| \mathbf{p}\right|$. Here we can also
assume that the electrons are confined to the LLL. Thus,
performing the same calculations as in (\ref{32})-(\ref{34}) we
arrive to a self-energy operator $\overline{\Sigma }(p,B)$ with a
structure identical to (\ref{31}), and coefficients given by an
expression similar to (\ref{34}), but with the Matsubara
replacement, $\int dq_{0}\rightarrow (2\pi i/\beta
)\sum\limits_{m=-\infty }^{\infty }(q_{4}=\frac{(2m+1)\pi }{\beta
},$ $m=0,\pm 1,\pm 2,...)$, corresponding to the discretization of
the fourth component $q_{4}$ of the fermion momentum, obtained
after introducing the Wick rotation to Euclidean space
$(q_{0}\rightarrow iq_{4})$, $\beta $ being the inverse of the
temperature $\beta =1/T$.

After summing in $m$ and extracting the vacuum part (\ref{34}), we
obtain that the coefficients corresponding to the thermal
contribution $\overline{\Sigma } _{T}(p,B)$ are given by
\[
a_{0}^{\prime }(T)=-c_{0}^{\prime }(T)=\frac{-i2\pi^{2}
g^{2}eB(p_{3}+ip_{4})}{ p_{\Vert }^{2}}e^{-p_{\bot
}^{2}/2eB}\delta ^{(4)}(p-p^{\prime })
\]
\begin{equation}
\int dq_{3}\left\{ \frac{q_{3}(p_{4}^{2}+\varepsilon
_{2}^{2}-\varepsilon _{1}^{2})+2ip_{4}\varepsilon
_{1}^{2}}{\varepsilon _{1}[(p_{4}^{2}+\varepsilon
_{2}^{2}-\varepsilon _{1}^{2})^{2}+4\varepsilon
_{1}^{2}p_{4}^{2}]}f_{F}(\varepsilon _{1})+\frac{(q_{3}-ip_{4})(p_{4}^{2}-%
\varepsilon _{2}^{2}+\varepsilon _{1}^{2})-2ip_{4}\varepsilon _{2}^{2}}{%
\varepsilon _{2}[(p_{4}^{2}-\varepsilon _{2}^{2}+\varepsilon
_{1}^{2})^{2}+4\varepsilon _{2}^{2}p_{4}^{2}]}f_{B}(\varepsilon
_{2})\right\} \label{40}
\end{equation}
where
\begin{equation}
f_{F}(\varepsilon _{1})=\frac{1}{1+e^{\beta \varepsilon
_{1}}},\qquad f_{B}(\varepsilon _{2})=\frac{1}{1-e^{\beta
\varepsilon _{2}}}  \label{41}
\end{equation}
are the fermion and boson distribution functions respectively,
with effective energies
\begin{equation}
\varepsilon _{1}=\sqrt{q_{3}^{2}+m_{e}^{2}},\qquad \varepsilon _{2}=\sqrt{%
(q_{3}-p_{3})^{2}+M_{W}^{2}-2eB}  \label{42}
\end{equation}

To do the integral in (\ref{40}) we take into account the approximation $%
M_{W}\gg \sqrt{B}\geq T\gg m_{e},\left| p\right| $. Thus, to
leading order we can neglect the contribution of the boson
distribution, since it is damped by the exponential factor
$e^{-M_{W}/T}$. Integrating in momentum and expanding in powers of
$1/M_{W}^{2}$, we obtain, up to leading order,
\begin{equation}
\overline{\Sigma }_{T}(p,B)=a_{0}(T)p\llap/_{\Vert }+c_{0}(T)p^{\mu }\widehat{%
\widetilde{F}}_{\mu \nu }\gamma ^{\nu }  \label{43}
\end{equation}
\begin{equation}
a_{0}(T)=-c_{0}(T)=\frac{-g^{2}eBT^{2}}{24M_{W}^{4}}e^{-p_{\bot
}^{2}/2eB} \label{44}
\end{equation}

Hence, the thermal contribution is one order smaller in powers of $%
1/M_{W}^{2}$ than the field-dependent vacuum contribution
(\ref{35}). Notice that in the strong-field approximation the
thermal contribution (\ref{44}) has the same $1/M_{W}^{4}$-order
as the corresponding thermal contributions at weak \cite{Weak-F}
and zero \cite{Raffelt} field.

We should call attention to the fact that when calculating the
thermal contribution to the e-W bubble by using the Schwinger
proper-time method \cite {Schwinger}, some authors \cite{Weak-F}
have considered the following expansion
\begin{equation}
G(k,B)_{\mu }\,_{\nu }\simeq \frac{\delta _{\mu \nu }}{M_{W}^{2}}+\frac{%
\delta _{\mu \nu }\Delta ^{2}-\Delta _{\mu }\Delta _{\nu }}{M_{W}^{4}}+%
\mathcal{O}(\frac{\mathnormal{\Delta ^{2}F_{\mu \nu
}}}{\mathnormal{M_{W}^{6}}}) \label{47}
\end{equation}
within the W-boson Green function in the presence of a constant
magnetic field
\begin{equation}
G(x,y)_{\mu }\,_{\nu }=\phi (x,y)\int \frac{d^{4}k}{(2\pi
)^{4}}e^{ik\cdot (x-y)}G(k,B)_{\mu }\,_{\nu },  \label{45}
\end{equation}
where
\begin{equation}
\phi (x,y)=\exp i\frac{e}{2}y_{\mu }F^{\mu \nu }x_{\nu }
\label{46}
\end{equation}
is the well-known phase factor depending on the applied field \cite{Dittrich}%
, and $\Delta (k)$ is the energy-momentum transfer.

Based on this expansion, it has been argued that the non-local
lower contribution to the bubble diagram should be of order
$1/M_{W}^{4}$, and that the local ($1/M_{W}^{2}$)-order term only
contributes at finite density. While this argument is correct when
applied to the statistical part of the self-energy operator (it
explains the $1/M_{W}^{4}$ order appearing in the thermal
coefficients $a_{0}(T)$ and $c_{0}(T)$ in (\ref{44})), it is not
valid for the self-energy vacuum contribution. For the vacuum
($T=0,$ $\mu =0$) contribution the situation is different, since
we cannot neglect the internal momenta as compared with $M_{W}$ in
the poles of the W-boson Green's function, due to the lack of the
ultraviolet cutoff which is present in statistics. Thus, when
calculating the vacuum part, the first term in (\ref{47}) should
be replaced by $\delta _{\mu \nu }/(M_{W}^{2}+k^{2})$. Such
momentum dependence in the pole of the W-boson Green's function
makes a nontrivial contribution of ($1/M_{W}^{2}$)-order into the
non-local part of the bubble diagram. Observe that the W-boson
Green's function obtained in our approach, Eq. (\ref{14}), only
has such structure indeed, and that the vacuum results at strong
(\ref {35}) and weak (\ref{36}) fields give rise to non-local
($1/M_{W}^{2}$)-order contributions to neutrino self-energy.

\subsection{Charged Medium in Strong Magnetic Field}\label{charged-medium}

We consider now a medium with a finite density of electric charge.
As usual, a finite density is reflected through the shift
$q_{0}\rightarrow q_{0}-\mu $ in the momenta of the particles with
non-zero charge associated to $\mu $. The chemical potential,
depending on the value of the charge density, plays the role of an
external parameter. In stellar medium, for example, the electric
charge is fixed by the net charge of the baryons, which, due to
their large masses, are usually treated as a classical background.
Here the following comment is in order, in electroweak matter at
finite density there exists the possibility to induce additional
``chemical potentials'' \cite{Shabad}. These chemical potentials
are nothing but dynamically generated background fields given by
the average $\left\langle W_{0}^{3}\right\rangle $ and
$\left\langle B_{0}\right\rangle $ of the zero components of the
gauge fields. They are known to appear, for instance, in the
presence of finite lepton/baryon density \cite{Shabad} (for recent
applications related to this effect see Refs.~\cite{Gusynin}). In
our case, a possible consequence of the condensation of such
average fields could be the modification of the effective chemical
potential of the W-boson, which then might be different from the
electron chemical potential. Nevertheless, that modification, even
if present in the case under study, will have not any relevant
consequence,
as the W-boson contribution will be always suppressed by the factor $%
e^{-M_{W}/T}$.

In the strong-field approximation (i.e. after considering the LLL
approximation in the neutrino self-energy (\ref{24-a})-(\ref{28}))
the non-zero coefficients of the neutrino self-energy in the
charged medium are given in Euclidean space by
\begin{equation}
a_{0}^{\prime }=-c_{0}^{\prime }=\frac{ig^{2}\pi^{2}
(p_{3}+ip_{4})}{\beta
p_{\Vert }^{2}}e^{-p_{\bot }^{2}/2eB}\sum\limits_{m}\int dq_{3}\frac{%
(q_{3}-iq_{4}^{*})}{[(q_{4}^{*})^{2}+\varepsilon
_{1}^{2}][(q_{4}^{*}-p_{4})^{2}+\varepsilon _{2}^{2}]}, \label{50}
\end{equation}
This expression is similar to Eq. (\ref{34}) after taking into
account the Matsubara replacement $\int dq_{0}\rightarrow (i/\beta
)\sum\limits_{m=-\infty }^{\infty }(q_{4}=\frac{(2m+1)\pi }{\beta },$ $%
m=0,\pm 1,\pm 2,...)$, with the additional change $%
q_{4}\rightarrow q_{4}^{*}$, where $q_{4}^{*}=q_{4}+i\mu $.

After carrying out the temperature sum in (\ref{50}), and
subtracting the vacuum part (\ref{34}), we arrive to
\[
a_{0}^{\prime }(T,\mu )=-c_{0}^{\prime }(T,\mu )=\frac{-i\pi^{2}
g^{2}eB(p_{3}+ip_{4})}{2p_{\Vert }^{2}}e^{-p_{\bot
}^{2}/2eB}\delta ^{(4)}(p-p^{\prime })\int dq_{3}
\]
\[
\times\{\frac{(q_{3}+\varepsilon _{1})}{\varepsilon
_{1}[(i\varepsilon
_{1}-p_{4})^{2}+\varepsilon _{2}^{2}]}f_{F}^{-}(\varepsilon _{1})+\frac{%
(q_{3}-\varepsilon _{1})}{\varepsilon _{1}[(i\varepsilon
_{1}+p_{4})^{2}+\varepsilon _{2}^{2}]}f_{F}^{+}(\varepsilon _{1})
\]
\begin{equation}
+\frac{(q_{3}+\varepsilon _{2}-ip_{4})}{\varepsilon
_{2}[(i\varepsilon
_{2}+p_{4})^{2}+\varepsilon _{1}^{2}]}f_{B}^{-}(\varepsilon _{2})+\frac{%
(q_{3}-\varepsilon _{2}-ip_{4})}{\varepsilon _{2}[(i\varepsilon
_{2}-p_{4})^{2}+\varepsilon _{1}^{2}]}f_{B}^{+}(\varepsilon
_{2})\} \label{52}
\end{equation}
where
\begin{equation}
f_{F}^{\mp }(\varepsilon _{1})=\frac{1}{1+e^{\beta (\varepsilon
_{1}\mp \mu )}},\qquad f_{B}^{\mp }(\varepsilon
_{2})=\frac{1}{1-e^{\beta (\varepsilon _{2}\mp \mu )}}  \label{53}
\end{equation}
are the fermion/anti-fermion and boson/anti-boson distributions
respectively. Here, we obtain an equal energy split for both
distributions since both charged particles have the same electric
charge and thus are characterized by the same chemical potential.

Assuming the approximation $M_{W}\gg \sqrt{B}\gg \mu ,T,m_{e},\left| \mathbf{%
p}\right| $, Eq. (\ref{52}) is reduced in leading order to
\begin{equation}
a_{0}^{\prime }(T,\mu )=-c_{0}^{\prime }(T,\mu )=\frac{-i\pi
^{4}g^{2}(p_{3}+ip_{4})}{M_{W}^{2}p_{\Vert }^{2}}[N_{0}^{-}-N_{0}^{+}%
]e^{-p_{\bot }^{2}/2eB}\delta ^{(4)}(p-p^{\prime })  \label{54}
\end{equation}
where the notation $N_{0}^{\mp }$ for the electron/positron number
densities in the LLL
\begin{equation}
N_{0}^{\mp }=\left| eB\right| \int \frac{dq_{3}}{2\pi
^{2}}f_{F}^{\mp }(\varepsilon _{1}).  \label{55}
\end{equation}
was introduced.

It is interesting to consider the situation where $\mu \gg T$,
which is a natural condition in many astrophysical environments.
In this case Eq. (\ref {54}) becomes
\begin{equation}
a_{0}^{\prime }(T,\mu )=-c_{0}^{\prime }(T,\mu )=\frac{-i2\pi^{2}
g^{2}eB(p_{3}+ip_{4})\mu }{M_{W}^{2}p_{\Vert }^{2}}e^{-p_{\bot
}^{2}/2eB}\delta ^{(4)}(p-p^{\prime })  \label{56}
\end{equation}

After analytic continuation to Minkowski space, we find that the
leading contribution to the statistical part of the self-energy in
a strongly magnetized charged medium is
\begin{equation}
\overline{\Sigma }_{T,\mu }(p,B)=a_{0}(T,\mu )p\llap/_{\Vert
}+c_{0}(T,\mu )p^{\mu }\widehat{\widetilde{F}}_{\mu \nu }\gamma
^{\nu }  \label{57}
\end{equation}
\begin{equation}
a_{0}(T,\mu )=-c_{0}(T,\mu )=\frac{g^{2}eB(p_{3}+p_{0})\mu
}{2(2\pi )^{2}M_{W}^{2}p_{\Vert }^{2}}e^{-p_{\bot }^{2}/2eB}
\label{58}
\end{equation}

The apparent dependence on the momentum components $p_{0}$,
$p_{3}$ of the coefficients $a_{0}(T,\mu )$, $c_{0}(T,\mu )$ is
deceiving, as we can easily rewrite Eqs. (\ref{57})-(\ref{58}) in
a more convenient way using the base formed by
the four-velocity $u_{\mu }$ and the covariant magnetic field vector $B_{\mu }=\frac{1}{2}%
\varepsilon _{\mu \nu \rho \lambda }u^{\nu }F^{\rho \lambda }$, as
\begin{equation}
\overline{\Sigma }_{T,\mu }(p,B)=\overline{a}_{0}(T,\mu )u\llap / +\overline{%
c}_{0}(T,\mu )B\llap/  \label{59}
\end{equation}
\begin{equation}
\overline{a}_{0}(T,\mu )=\frac{-g^{2}eB\mu }{2(2\pi )^{2}M_{W}^{2}}%
e^{-p_{\bot }^{2}/2eB},\qquad \overline{c}_{0}(T,\mu )=\frac{g^{2}e\mu }{%
2(2\pi )^{2}M_{W}^{2}}e^{-p_{\bot }^{2}/2eB}  \label{60}
\end{equation}

$ $From (\ref{59})-(\ref{60}) we see that the leading term in the
statistical part of the self-energy in the charged medium in
strong magnetic field is independent of the momentum and
proportional to $1/M_{W}^{2}$, thus of larger order than the
statistical contribution of the neutral case (Eq. (\ref {44})).

Comparing our result (\ref{59})-(\ref{60}) with those found at
weak field \cite{Nieves}, we see that the coefficient of the
structure $B\llap /$ has similar qualitative characteristics. That
is, both $c$ coefficients are linear in $\mu $ and have the same
order, $1/M_{W}^{2}$. Nevertheless, the coefficients of the
structure $u\llap / $ significantly differ in the strong and weak
cases. In the weak-field limit $\overline{a}$ only depends on the
density, thus its leading contribution is proportional to $\mu
^{3}$ (this is the characteristic term of the MSW effect
\cite{msw}). However, for fields larger than $\mu $, the relation
$\mu B\gg \mu ^{3}$ holds, and the leading becomes proportional to
$\mu B$, as in Eq.~(\ref{60}). Thus, in the strong field case it
is the field the parameter that drives the main dependence of both
structures in $\overline{\Sigma }_{T,\mu }$.

\section{Neutrino Index of Refraction in Strong Magnetic Field}\label{index-ref}

The neutrino index of refraction, defined by
\begin{equation}
n=\frac{\mid \mathbf{p}\mid }{E(\mid \mathbf{p}\mid )}  \label{61}
\end{equation}
where $\mathbf{p}$ and $E$ are the neutrino momentum and energy
respectively, is a quantity that plays a significant role in
neutrino flavor oscillations in a medium \cite{msw}. A
flavor-dependent index of refraction would enhance the periodical
change between different flavors of neutrinos travelling through
the medium.

To find the index of refraction we need to know the dispersion relation $%
E=E(\mid \mathbf{p}\mid )$. In the presence of a medium the energy
spectra of the massless left-handed neutrinos are modified due to
their weak interaction with the particles of the background. In a
magnetized medium, even though the neutrinos are electrically
neutral, they feel the magnetic field indirectly through their
interaction with the charged particles of the medium whose
properties are affected by the field.

As discussed in Sec.\ref{selfen-gen-struc}, in a magnetized medium
the general form of the neutrino dispersion relation is given by
Eq. (\ref{5}), with the self-energy depending on the medium
characteristics. As follows, we present the neutrino energy
spectrum and index of refraction for magnetized neutral and
charged media.

\subsection{Neutrino Index of Refraction in Strongly Magnetized Neutral Medium}\label{index-ref-neut-med}

In the strongly magnetized neutral medium the dispersion relation
is obtained from
\begin{equation}
\det \left[ p\llap/+\overline{\Sigma }_{0}(p,B)+\overline{\Sigma }%
_{T}(p,B)\right] =0,  \label{62}
\end{equation}
with $\overline{\Sigma }_{0}(p,B)$ and $\overline{\Sigma
}_{T}(p,B)$ explicitly given in Eqs. (\ref{34-a})-(\ref{35}) and
(\ref{43})-(\ref{44}) respectively. The solution of (\ref{62}) in
leading order in the Fermi coupling constant is
\begin{equation}
p_{0}=\pm
\left| \mathbf{p} +\sqrt{2/a^{(0)}}(\mathbf{p}\times \widehat{\mathbf{B}}%
)\right| \simeq \pm \left| \mathbf{p}\right| [1-a^{(0)}\sin
^{2}\alpha] \label{63}
\end{equation}
In (\ref{63})$,  \alpha $ is the angle between the direction of
the neutrino momentum and that of the applied magnetic field, and
the coefficient $a^{(0)}=a_{0}+a_{0}(T)$ is explicitly given by

\begin{equation}
a^{(0)}=\frac{-g^{2}eB}{8M_{W}^{2}}[\frac{1}{(2\pi)^{2} }+\frac{T^{2}}{%
3M_{W}^{2}}]\exp (-p_{\perp }^{2}/eB),  \label{64}
\end{equation}

$ $From (\ref{64}) it is clear that in the strong-field limit the
thermal correction $a_{0}(T)$ to the neutrino energy is much
smaller than the vacuum correction $a_{0}$. As we will discuss in
Sec.\ref{conclusions}, this result can be significant for
cosmological applications in case that a strong primordial
magnetic field could exist in the primeval plasma.

The neutrino ($+$)/anti-neutrino ($-$) energies are given by
 $E_{\pm }\equiv \pm p_{0}$ respectively. Substituting Eq. (\ref{63}) into (\ref{61}) we obtain for the
neutrino (antineutrino) index of refraction in a neutral medium
\begin{equation}
n\simeq 1+a^{(0)}\sin ^{2}\alpha  \label{65}
\end{equation}

$ $Eq. (\ref{65}) implies that the index of refraction depends on
the neutrino direction of motion. The order of the anisotropy is
$g^{2}\frac{\left| eB\right| }{M_{W}^{2}}$. Maximum field effects
take place for neutrinos propagating perpendicularly to the
magnetic field. That is, the maximum departure of the neutrino
phase velocity from the light velocity $c$, occurs at $\alpha =\pi
/2$. Notice that in the neutral magnetized medium the magnetic
field effect does not differentiate between neutrinos and
antineutrinos.

Here the following comment is in order. If we consider the
weak-field results (\ref{36}) obtained in Ref.~\cite{McKeon} for
the neutrino self-energy, in the dispersion relation (\ref{6}) we
obtain
\begin{equation}
E_{\pm}^{\prime }= \left| \mathbf{p}\right| [1+\frac{5}{18}\widetilde{c}%
_{0}^{2}\sin ^{2}\alpha ]  \label{66}
\end{equation}
In (\ref{66}), as the energy depends on $\widetilde{c}_{0}^{2}$,
we can see that the weak field produces a negligible second order
correction in term of the Fermi coupling constant expansion (i.e.
a $1/M_{W}^{4}$-order effect). It can be corroborated that the
inclusion of temperature in this approximation also produces a
second order correction \cite{Weak-F}.

Thus, we conclude that the strong field produces an effect
qualitatively larger, which is even more important than the
thermal one. We call reader's attention to the fact that the
field-dependent vacuum contribution to the neutrino energy
(\ref{63}) has the same order in the Fermi coupling constant as
the ones found in a charged medium at zero \cite{msw} and weak
\cite{Nieves} fields.

\subsection{Neutrino Index of Refraction in Strongly Magnetized Charged Medium}\label{ind-ref-charged-medium}

Now we consider the effect of a strong magnetic field in neutrino
propagation in a charged medium. In this case the dispersion
relation is
\begin{equation}
\det \left[ p\llap/+\overline{\Sigma }_{0}(p,B)+\overline{\Sigma
}_{T,\mu }(p,B)\right] =0,  \label{67}
\end{equation}
with $\overline{\Sigma }_{0}(p,B)$ and $\overline{\Sigma }_{T,\mu
}(p,B)$ given in Eqs. (\ref{34-a})-(\ref{35}) and
(\ref{59})-(\ref{60}) respectively. The solution of (\ref{67}) is

\begin{equation}
E_{\pm }=\pm \frac{4a_{0}\mu}{(1-2a_{0})}+\sqrt{\frac{\left|
\mathbf{p}\right| ^{2}-2a_{0}[p_{3}-4\mu]p_{3}}{1-2a_{0}}%
}.  \label{67-a}
\end{equation}

In leading order in the Fermi coupling constant, expression
(\ref{67-a}) is approximated by
\begin{equation}
E_{\pm }\simeq \left| \mathbf{p}\right| [1-a_{0}\sin ^{2}\alpha
]-\mathbf{\mathcal{M}.B}\pm \mathnormal{E_{0}} \label{68}
\end{equation}
where
\begin{equation}
\mathcal{M} \equiv \frac{-\mathnormal{E_{0}}}{\left|
\mathbf{B}\right|}\frac{\textbf{p}}{\left|
\mathbf{p}\right|}\label{69},
\end{equation}
and
\begin{equation}
E_{0}=\sqrt{2}G_{F}e^{-p_{\bot }^{2}/2eB}[N_{0}^{-}-N_{0}^{+}]
\label{70}
\end{equation}

In the RHS of Eq. (\ref{68}), the first term is the modified
neutrino kinetic energy, with the field-dependent radiative
correction already found in the neutral case (see Eq. (\ref{63}));
the second term can be interpreted as a
magnetic-moment/magnetic-field interaction energy, with
$\mathcal{M}$ playing the role of a neutrino effective magnetic
moment; and the third term is a rest energy induced by the
magnetized charged medium.

We underline that, contrary to the case of the neutrino anomalous
magnetic moment found beyond the standard model by including
right-chiral neutrinos \cite{Fujikawa}, the induced magnetic
moment here obtained does not require a massive neutrino. We
remind that we are only considering left-chiral Dirac neutrinos,
so the magnetic moment in (\ref{68}) cannot be associated with the
anomalous magnetic moment structural term $\sigma ^{\mu \nu
}F_{\mu \nu } $ in the self-energy. This structure is forbidden in
the present case by the explicit chirality of the standard model.
In the case under study we have that although the neutrino is a
neutral massless particle, the charged medium can endow it,
through quantum corrections, with a magnetic moment proportional
to the induced neutrino rest energy. Thus, we are in the presence
of a peculiar magnetic moment created thanks to the
particle-antiparticle unbalance of the charged medium. Such a
charge asymmetry permits the formation of a net virtual current
that, due to the magnetic field, circulates in a plane
perpendicular to the neutrino propagation and therefore produces
an effective magnetic moment in the direction of the neutrino
momentum.

In a charged medium CP-symmetry is violated \cite{Pal}, and a
common property of electroweak media with CP violation is a
separation between neutrino and anti-neutrino energies. In the
present case this property is manifested in the double sign of
$E_{0}$ in (\ref{68}). Although this difference depends on the
magnetic field entering in $N_{0}^{\pm }$, we stress that it
depends on the magnitude of the magnetic field but not on its
direction. Thus, the energy anisotropy connected to the first two terms of (%
\ref{68}) is the same for neutrinos and anti-neutrinos.

The neutrino/antineutrino energy in the charged medium in the
presence of a weak field has been found \cite{Nieves} to be
\begin{equation}
E_{\pm }^{\prime \prime }=\left| \mathbf{p}\right| -b^{\prime }\frac{\mathbf{%
p}\cdot \mathbf{B}}{\left| \mathbf{p}\right| }\pm \,a^{\prime },
\label{71}
\end{equation}
with
\begin{equation}
a^{\prime }=\frac{g^{2}}{4M_{W}^{2}}(N_{-}-N_{+}),\qquad b^{\prime }=\frac{%
eg^{2}}{2M_{W}^{2}}\int \frac{d^{3}p}{(2\pi )^{3}2E}\frac{d}{dE}%
(f_{-}-f_{+}),  \label{72}
\end{equation}
where $N_{\pm }$ are the electron/positron number densities, and
$f_{\pm }$ the electron/positron distributions.

Comparing the strong-field (\ref{68}) and the weak-field
(\ref{71}) dispersion relations, we see that the field-dependent
correction to the kinetic energy appearing in the strong-field
approximation is absent in the leading order of the weak-field
case. For a strong-field, that correction turns out to be
important for neutrinos
propagating perpendicular to the magnetic field. Another difference between (%
\ref{68}) and (\ref{71}), is that in the last dispersion relation
the rest energy $a^{\prime }$ does not depend on the magnetic
field. This is the pure
medium contribution that gives rise to the well known MSW effect \cite{msw}%
. As discussed above, in the strong-field case, where all
parameters,
including $\mu $ and $\left| \mathbf{p}\right| $ are assumed smaller than $%
\sqrt{eB}$, such a term is negligible compared to the contribution
$E_{0}$ which is proportional to the field. In both approximations
there is an additional anisotropy related to the induced magnetic
moment term, which is linear in the magnetic field and depends on
the unbalance between particles and anti-particles. We should
notice that this anisotropic term changes its sign when the
neutrino reverses its motion. This property is crucial for a
possible explanation of the peculiar high pulsar velocities
\cite{Kusenko}. Nevertheless, for the anisotropic term that
modifies the kinetic energy in the strong-field approximation, the
neutrino
energy-momentum relation is invariant under the change of $\alpha $ by $%
-\alpha $.

Substituting (\ref{68}) into (\ref{61}) we obtain that the
neutrino/anti-neutrino index of refraction in the charged medium
is given by
\begin{equation}
n_{\pm}=1+a_{0}\sin ^{2}\alpha +\frac{\mathcal{M\cdot }\mathbf{B}\mp\mathnormal{E_{0}}}{%
\left| \mathbf{p}\right| }  \label{73}
\end{equation}
The result (\ref{73}) implies that the index of refraction depends
on the neutrino direction of motion relative to the magnetic
field. Moreover, neutrinos and antineutrinos have different index
of refractions and therefore different phase velocities even if
they move in the same direction. For neutrinos, if
$N_{0}^{-}>N_{0}^{+}$ (i.e. if there is a larger number of
electrons than positrons) the index of refraction for $B\neq0$ is
smaller than one, so their phase velocities will be smaller than
light velocity. Thus, in the charged medium with strong magnetic
field, neutrinos behave as massive particles. This is in agreement
with the behavior they have in a dense medium, even in the absence
of magnetic field \cite{msw}. On the other hand, for
anti-neutrinos the index of refraction (the one with the positive
sign in front of $E_{0}$ in (\ref{73})) can be larger than one.
For instance, for anti-neutrinos moving opposite to the field-line
directions the index of refraction is
\begin{equation}
n_{-}=1+\frac{2E_{0}}{\left| \mathbf{p}\right| }>1  \label{74}
\end{equation}

Thus, antineutrinos will have phase velocities larger than $c$ and
depending on the magnetic field strength and electron density. In
this regard we should mention that particles with zero rest mass
can even have group velocities that exceed $c$ in anomalously
dispersive media \cite{velocities}. Moreover, in non-trivial
backgrounds, particles with superluminal propagations are not
intrinsically forbidden in quantum field theories. For instance,
superluminal photons appears in curved spaces \cite{Drummond},
Casimir vacua \cite{Scharnhorst} and in QED with compactified
spatial dimensions \cite{Romero}. Discussions about the
non-violation of microcausality by the existence of such
superluminal velocities, as well as the lack of a contradiction
between such a phenomenon with the bases of special theory of
relativity, can be found in Refs.~ \cite{velocities},\cite
{Drummond},\cite{Shore}. There, it was noted that the ``front
velocity,'' the one related to the index of refraction in the
infinite frequency limit, is the one that cannot exceed $c,$ since
it is related to the signal transmission.

\section{Strong Field and LLL Approximation. Numerical Test}\label{numerical}

In the calculation of the neutrino self-energy in the strong-field limit ($%
M_{W}\gg \sqrt{B}\gg m_{e},T,\mu ,\left| \mathbf{p}\right| $) done
in Sec.~\ref{strong-b-selfen}, we assumed that the main
contribution to the self-energy loop diagram came from electrons
in the LLL, since in the strong-field limit the energy gap between
electron Landau levels is much larger than the electron average
energy in the medium. In this Section we will check the validity
of these arguments with the help of numerical calculations.

Taking into account that the relevant physical quantity in this
study is the neutrino energy, we can concentrate our analysis on
the self-energy coefficients that give the largest contribution to
the dispersion relation. It is easy to check, from the analytical
structure of Eqs. (\ref{25})-(\ref {28}), that the leading term in
the expansion of the coefficients in powers of $1/M_{W}^{2}$ can
be at most of order $1/M_{W}^{2}$. Keeping in mind that the only
coefficients that could contribute with a $1/M_{W}^{2}$-order term
in the dispersion relation (\ref{6}) are $a$ and $b$ (coefficients $c$ and $%
d $ appear squared in (\ref{6}) and their contribution to the
dispersion relation should start at least with terms of order
$1/M_{W}^{4}$), we can approximate the dispersion relation
(\ref{6}) by
\begin{equation}
p_{0}^{2}\simeq \left| \mathbf{p}\right| ^{2}-2(a-b)p_{\bot }^{2}
\label{74-a}
\end{equation}

Then, to validate the LLL approximation we should numerically
study the ratio
\begin{equation}
\varkappa =\frac{a-b}{a_{0}-b_{0}},  \label{74-b}
\end{equation}
where $a$ and $b$ are respectively obtained from Eqs. (\ref{25})
and (\ref {26}), summing in all the Landau levels and taking
parameter values in
the strong-field region $M_{W}\gg \sqrt{B}\gg m_{e},$ $T,$ $\left| \mathbf{%
p}\right| $. The coefficients $a_{0}$ and $b_{0}$ are the
corresponding values at $T=0$ in the LLL . In the denominator of
(\ref{74-a}) we are neglecting the thermal contributions
$a_{0}(T)$ and $b_{0}(T)$ since they would make an insignificant
contribution ($1/M_{W}^{2}$-order smaller) as compared with the
vacuum ones in the parameter range $T \ll M_{W}$. In
Sec.~\ref{strong-b-selfen}, we showed that $b_{0}=0$, and that
$a_{0}$ is given by Eq. (\ref{35}).

The strong-field conditions can be naturally found in many
astrophysical systems like magnetars, neutron stars, etc. In
cosmological applications, however, the viable primordial fields
can never be much larger
than the temperature\footnote{%
In cosmology the electric chemical potential is so small that we
can always assume it is zero.}, as according to the equipartition
principle, the magnetic energy can only be a small fraction of the
universe energy density. This argument leads to the relation
$eB/T^{2}\sim \mathcal{O}(\emph{1})$. Clearly, this is not a
situation very consistent with a strong-field approximation.
However, even under this condition, it is natural to expect that
the thermal energy should be barely enough to induce the
occupation of just a few of the lower electron Landau levels, and
one would not be surprised if the LLL approximation still gives
the leading contribution to the dispersion relation.

To investigate (\ref{74-b}) it will be more convenient to
independently study the variation range for each coefficient
ratio,
\begin{equation}
\varkappa _{a}=\frac{a(T=0)+a(T)}{a_{0}}-\varkappa
_{a}(B=0,T=0),\qquad \varkappa
_{b}=\frac{b(T=0)+b(T)}{a_{0}}-\varkappa _{b}(B=0,T=0),
\label{74-c}
\end{equation}
where $a(T=0)$ ($b(T=0)$) and $a(T)$ ($b(T)$) are respectively the
vacuum and thermal contributions of each coefficient $a$ ($b$).
Here we subtract the ultraviolet divergent zero-field,
zero-temperature parts $\varkappa _{a}(B=0,T=0),$ $\varkappa
_{b}(B=0,T=0),$ as they can only contribute, after
renormalization, with negligible zero-field terms. The validity of
the LLL approximation should imply that $\varkappa _{a}\approx 1$
and $\varkappa _{b}\ll 1$ for the range of parameters considered.
In the first part of this Section we examine the
validity of the LLL in the zero-temperature case obtaining $%
\varkappa _{a}(T=0)=a(T=0)/a_{0}-\varkappa _{a}(B=0,T=0)\approx
1$, $\varkappa _{b}(T=0)=b(T=0)/a_{0}-\varkappa _{b}(B=0,T=0)\ll
1$. In the second part, we do a similar analysis, for the thermal
part of $a$, finding that $\varkappa _{a}(T)=a(T)/a_{0}\ll 1$, as
expected from the theoretical considerations given in
Sec.~\ref{strong-b-selfen} (see the discussion below Eq.
(\ref{46})). As this last numerical calculation is performed for
magnetic fields that can even be a few orders smaller than
$T^{2},$ the result $\varkappa _{a}(T)\ll 1$ confirms the
appropriateness of the LLL-approximation for a rather wide range
of primordial fields. The analysis of the finite temperature part
of coefficient $b$ ($\varkappa _{b}(T)=b(T)/a_{0}$) is not
explicitly done in the paper, however, it is not hard to see that
it gives rise to a similar result, that is, $\varkappa _{b}(T)\ll
1$.

\subsection{Vacuum Contribution}\label{numerical-vac}

If we introduce in expressions (\ref{25}) and (\ref{26}) the
integral representations
\begin{equation}
\frac{1}{\overline{q}^{2}+m_{e}^{2}}=\int\limits_{0}^{\infty }d\alpha \exp -(%
\overline{q}^{2}+m_{e}^{2})\alpha ,  \label{78}
\end{equation}
\begin{equation}
\frac{1}{\overline{k}^{2}+M_{W}^{2}}=\int\limits_{0}^{\infty }d\beta \exp -(%
\overline{k}^{2}+M_{W}^{2})\beta ,  \label{79}
\end{equation}
and take into account the recurrence relation \cite{Rainville}
\begin{equation}
(2l)^{1/2}\varphi _{l-1}(\xi )=(\partial _{\xi }+\xi )\varphi
_{l}(\xi ), \label{80}
\end{equation}
in (\ref{26}), it is possible to rewrite coefficients $a^{\prime }$ and $%
b^{\prime }$ as
\[
a^{\prime }=\frac{-ig^{2}eB}{2\left( 2\pi \right) ^{6}p_{\Vert
}^{2}}\int d^{4}x\int d^{4}y\int d^{3}\widehat{q}\int
d^{3}\widehat{k}\quad
e^{-i(p_{1}x_{1}-p_{1}^{\prime }.y_{1})}e^{i\widehat{x}\cdot (\widehat{k}+%
\widehat{q}-\widehat{p})+i\widehat{y}\cdot (\widehat{p}^{\prime }-\widehat{k}%
-\widehat{q})}
\]
\begin{equation}
\int\limits_{0}^{\infty }d\alpha \int\limits_{0}^{\infty }d\beta
\{e^{-(q_{\Vert }^{2}+m_{e}^{2})\alpha -(k_{\Vert
}^{2}+M_{W}^{2}-eB)\beta
}[(q_{0}-q_{3})(p_{3}+p_{0})-(q_{0}+q_{3})(p_{3}-p_{0})r^{2}t]S_{\alpha
}S_{\beta }\}  \label{75}
\end{equation}
\[
b^{\prime }=\frac{ig^{2}(eB)^{3/2}}{\left( 2\pi \right) ^{6}p_{\bot }^{2}}%
\int d^{4}x\int d^{4}y\int d^{3}\widehat{q}\int
d^{3}\widehat{k}\quad
e^{-i(p_{1}x_{1}-p_{1}^{\prime }.y_{1})}e^{i\widehat{x}\cdot (\widehat{k}+%
\widehat{q}-\widehat{p})+i\widehat{y}\cdot (\widehat{p}^{\prime }-\widehat{k}%
-\widehat{q})}
\]
\begin{equation}
\int\limits_{0}^{\infty }d\alpha \int\limits_{0}^{\infty }d\beta
\{e^{-(q_{\Vert }^{2}+m_{e}^{2})\alpha -(k_{\Vert }^{2}+M_{W}^{2}-eB)\beta }[%
\frac{\rho +\rho ^{\prime }}{1+t}p_{2}+i\frac{\rho -\rho ^{\prime }}{1-t}%
p_{1}]rtS_{\alpha }S_{\beta }\}  \label{76}
\end{equation}
Here, $S_{\alpha }$ and $S_{\beta }$ are found doing the sum in
Landau levels with the help of Mehler's formula \cite{Bateman}
\begin{equation}
S_{\alpha }=\sum\limits_{l=0}^{\infty }\varphi _{l}(\rho )\varphi
_{l}(\rho ^{\prime })t^{l}=[\pi (1-t^{2})]^{-1/2}\exp -(\frac{\rho
^{2}-\rho ^{\prime 2}}{2})\exp -\frac{(\rho ^{\prime }-\rho
t)^{2}}{1-t^{2}}  \label{77}
\end{equation}
where $t=\exp -(2eB\alpha )$, $\rho =\sqrt{2eB}(x_{1}+q_{2}/eB)$
and $\rho
^{\prime }=\sqrt{2eB}(y_{1}+q_{2}/eB).$ $S_{\beta }$ is obtained from Eq. (%
\ref{77}) by replacing $t\rightarrow r=\exp -(2eB\beta )$, $\rho
\rightarrow \xi =\sqrt{2eB}(x_{1}-k_{2}/eB)$ and $\rho ^{\prime
}\rightarrow \xi ^{\prime }=\sqrt{2eB}(y_{1}-k_{2}/eB)$ .

Integrating in $\widehat{x}$, $\widehat{y}$, $\widehat{q}$ and
$k_{2}$, and introducing the variable changes $x_1\rightarrow \xi$
and $y_1\rightarrow \xi^{\prime}$, we find
\[
a^{\prime }=\frac{-ig^{2}eB\pi }{p_{\Vert }^{2}}\delta
^{4}(p-p^{\prime })\int\limits_{-\infty }^{\infty }d^{2}k_{\Vert
}\int\limits_{0}^{\infty }d\alpha \int\limits_{0}^{\infty }d\beta
\int\limits_{-\infty }^{\infty }d\xi \int\limits_{-\infty
}^{\infty }d\xi ^{\prime }\{e^{-[(q_{\Vert }-p_{\Vert
})^{2}+m_{e}^{2}]\alpha -(k_{\Vert }^{2}+M_{W}^{2}-eB)\beta
}e^{-i(\xi -\xi ^{\prime })p_{1}/\sqrt{eB}}
\]
\begin{equation}
\lbrack
(1+r^{2}t)(p_{0}^{2}-p_{3}^{2})-(k_{0}-k_{3})(p_{3}+p_{0})+(k_{0}+k_{3})(p_{3}-p_{0})r^{2}t]S_{\alpha
}^{\prime }S_{\beta }\} \label{81}
\end{equation}
\[
b^{\prime }=\frac{2ig^{2}(eB)^{3/2}\pi }{p_{\bot }^{2}}\delta
^{4}(p-p^{\prime })\int\limits_{-\infty }^{\infty }d^{2}k_{\Vert
}\int\limits_{0}^{\infty }d\alpha \int\limits_{0}^{\infty }d\beta
\int\limits_{-\infty }^{\infty }d\xi \int\limits_{-\infty
}^{\infty }d\xi ^{\prime }\{e^{-[(q_{\Vert }-p_{\Vert
})^{2}+m_{e}^{2}]\alpha -(k_{\Vert }^{2}+M_{W}^{2}-eB)\beta }
\]
\begin{equation}
\lbrack \frac{\xi +\xi ^{\prime }}{1+t}p_{2}+i\frac{\xi -\xi ^{\prime }}{1-t}%
p_{1}+\frac{2p_{2}^{2}}{\sqrt{eB}(1+t)}]e^{-i(\xi -\xi ^{\prime })p_{1}/%
\sqrt{eB}}rtS_{\alpha }^{\prime }S_{\beta }\}  \label{82}
\end{equation}
where
\[
S_{\alpha }^{\prime }=[\pi (1-t^{2})]^{-1/2}\exp -(\frac{[\xi +p_{2}/\sqrt{eB%
}]^{2}-[\xi ^{\prime }+p_{2}/\sqrt{eB}]^{2}}{2})
\]
\begin{equation}
\exp -\frac{[(\xi ^{\prime }+p_{2}/\sqrt{eB})-(\xi +p_{2}/\sqrt{eB})t]^{2}}{%
1-t^{2}}  \label{83}
\end{equation}

\begin{figure}
\includegraphics{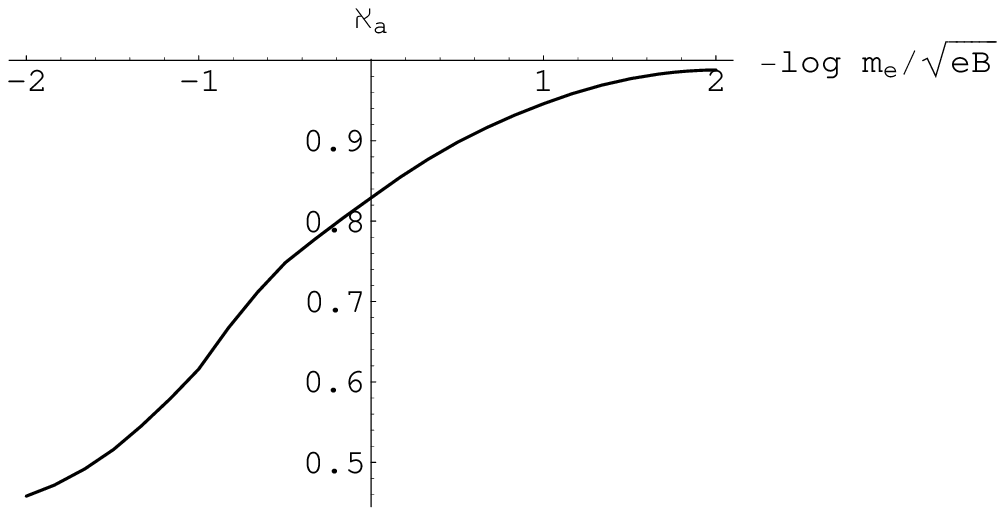}
\caption{\label{a-coefT0}Plot of $\varkappa _{a}(T=0)$ vs. $\log
\widehat{m}_{e}^{-1}$ for field range $10^{-2}m_{e}\leq
\sqrt{eB}\leq 10^{2}m_{e}$ and parameter values
$|\widehat{p}_{\parallel }|=|\widehat{p}_{\perp }|=10^{-1}
\sqrt{2} \widehat{m}_{e}$, $\widehat{M}_{W}=10^{5}
\widehat{m}_{e}$}
\end{figure}

\begin{figure}
\includegraphics{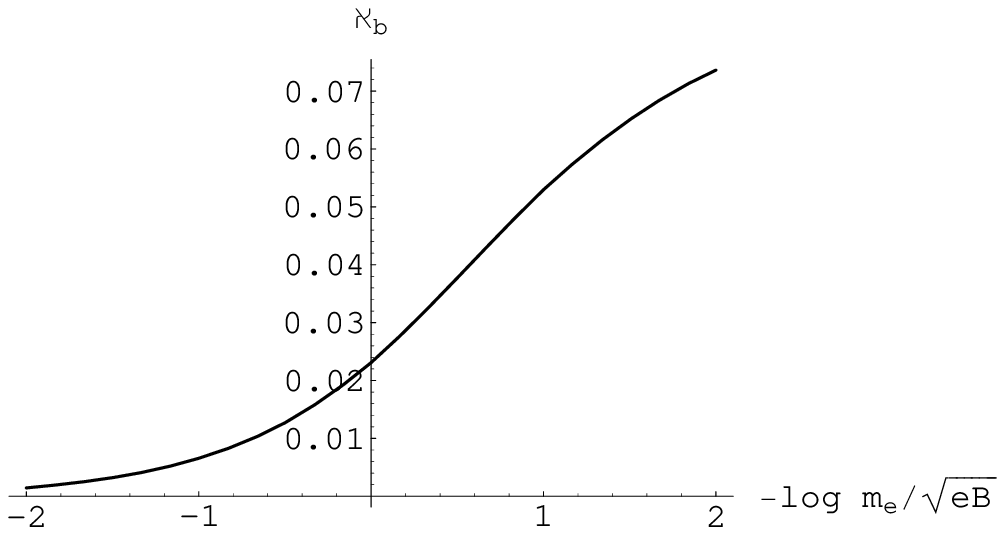}\caption{\label{b-coefT0}Plot of $\varkappa _{b}(T=0)$ vs. $\log
\widehat{m}_{e}^{-1}$ for field range $10^{-2}m_{e}\leq
\sqrt{eB}\leq 10^{2}m_{e}$ and parameter values
$|\widehat{p}_{\parallel }|=|\widehat{p}_{\perp }|=10^{-1}
\sqrt{2} \widehat{m}_{e}$, $\widehat{M}_{W}=10^{5}
\widehat{m}_{e}$}
\end{figure}

After Wick rotating to Euclidean space and doing the Gaussian integrals in $%
k_{3}$, $k_{4}$, $\xi $ and $\xi ^{\prime }$, we finally obtain
\[
a^{\prime }=\delta ^{4}(p-p^{\prime })g^{2}\pi
^{2}\int\limits_{0}^{\infty
}d\alpha \int\limits_{0}^{\infty }d\beta \frac{\beta }{(\alpha +\beta )^{2}}%
\frac{\cosh (\alpha +2\beta )}{\sinh (\alpha +\beta )}\exp (-\widehat{m}%
_{e}^{2}\alpha -\widehat{M}_{W}^{2}\,\beta )
\]
\begin{equation}
\exp -[\frac{\alpha \beta }{\alpha +\beta }\widehat{p}_{\Vert }^{2}+\frac{1}{%
cth(\alpha )+cth(\beta )}\widehat{p}_{\bot }^{2}]  \label{84}
\end{equation}
\[
b^{\prime }=-\delta ^{4}(p-p^{\prime })g^{2}\pi
^{2}\int\limits_{0}^{\infty
}d\alpha \int\limits_{0}^{\infty }d\beta \frac{1}{(\alpha +\beta )}\frac{%
\sinh (\beta )}{\sinh ^{2}(\alpha +\beta )}\exp (-\widehat{m}_{e}^{2}\alpha -%
\widehat{M}_{W}^{2}\,\beta )
\]
\begin{equation}
\exp -[\frac{\alpha \beta }{\alpha +\beta }\widehat{p}_{\Vert }^{2}+\frac{1}{%
cth(\alpha )+cth(\beta )}\widehat{p}_{\bot }^{2}]  \label{85}
\end{equation}
where we introduced the normalized parameters $\widehat{m}_{e}=m_{e}/\sqrt{eB%
}$, $\widehat{p}_{\mu }=p_{\mu }/\sqrt{eB}$, and $\widehat{M}_{W}=M_{W}/%
\sqrt{eB}$. Eqs. (\ref{84}) and (\ref{85}) are as far as we can go
in the calculation of $a^{\prime }$ and $b^{\prime }$ after
summing in all Landau levels without using any approximation.

In the zero-field limit the coefficients (\ref{84}) and (\ref{85})
reduce to
\begin{equation}
a_{B=0}^{\prime }=b_{B=0}^{\prime }=-\delta ^{4}(p-p^{\prime
})g^{2}\pi
^{2}\int\limits_{0}^{\infty }d\alpha \int\limits_{0}^{\infty }d\beta \frac{%
\beta }{(\alpha +\beta )^{3}}\exp -(\alpha +\frac{\alpha \,\beta
}{\alpha +\beta }\,\overline{p}_{\mu
}^{2}+\overline{M}_{W}^{2}\beta )  \label{86}
\end{equation}
where $\overline{M}_{W}=M_{W}/m_{e},$ $\overline{p}_{\mu }=p_{\mu
}/m_{e}$.

Our goal now is to investigate the validity of the LLL
approximation for the zero-temperature contribution of the
coefficients in the parameters' range: $M_{W}\gg \sqrt{B}\gg
m_{e},\left| \mathbf{p}\right| $. With this purpose, in Figs. 1
and 2 we plot,
\begin{equation}
\varkappa _{a}(T=0)=\frac{a-a_{B=0}}{a_{0}}\qquad \text{vs}\qquad
\log \widehat{m} ^{-1}\in \text{ }[-2,2]  \label{87}
\end{equation}
and
\begin{equation}
\varkappa _{b}(T=0)=\frac{b-b_{B=0}}{a_{0}}\qquad \text{vs}\qquad
\log \widehat{m} ^{-1}\in \text{ }[-2,2]  \label{88}
\end{equation}
respectively. Here $a_{0}$ is the LLL coefficient defined in Eq.
(\ref{35}). Notice that primed and unprimed variables are related
through: $a^{\prime }=\left( 2\pi \right) ^{4}\delta
^{(4)}(p-p^{\prime })\,a,$ $b^{\prime }=\left( 2\pi \right)
^{4}\delta ^{(4)}(p-p^{\prime })\,b$.

$ $From Fig. 1 it can be seen that $\varkappa _{a}(T=0)\approx 1$
when $\sqrt{eB} \geq m_{e},\left| \mathbf{p}\right|$ , while, from
\ Fig. 2, we see that $ \varkappa _{b}(T=0)\ll 1$ in all the range
$10^{-2}m_{e}\leq \sqrt{eB}\leq 10^{2}m_{e}$. This result
corroborates the initial assumption that the LLL approximation
gives the leading contribution to the vacuum part of the neutrino
self-energy at strong fields ($M_{W}\gg \sqrt{B}\gg m_{e},\left|
p\right| $).

\subsection{Thermal Contribution}\label{numerical-therm}

Let us consider now the finite temperature contributions. The
finite temperature part of the coefficient $a^{\prime }$ can be
found in the following way. We start from Eq. (\ref{81}), perform
the Wick rotation to Euclidean space, replace the integral in the
four momentum by the temperature sum $\int
dk_{4}\rightarrow (2\pi T)\sum\limits_{n=-\infty }^{\infty }$ , (with $%
k_{4}=2\pi nT),$ and integrate in $k_{3\text{ }},$ to obtain

\[
\hspace*{-4mm}
a^{\prime }=\delta ^{4}(p-p^{\prime })\frac{g^{2}eB\pi \sqrt{\pi }}{%
p_{\Vert }^{2}}\int\limits_{0}^{\infty }d\alpha
\int\limits_{0}^{\infty }d\beta \int\limits_{-\infty }^{\infty
}d\xi \int\limits_{-\infty }^{\infty }d\xi ^{\prime }\{S_{\alpha
}^{\prime }S_{\beta }\frac{e^{\alpha ^{2}p_{3}^{2}/(\alpha +\beta
)}}{\sqrt{\alpha +\beta }}e^{-[(p_{\Vert })^{2}+m_{e}^{2}]\alpha
-(M_{W}^{2}-eB)\beta }e^{-i(\xi -\xi ^{\prime })p_{1}/\sqrt{eB}}
\]
\[
\times (2\pi T)\sum\limits_{n=-\infty }^{\infty
}\{(1+r^{2}t)(p_{0}^{2}-p_{3}^{2})+\frac{\alpha p_{3}}{\alpha +\beta }[%
(p_{3}+p_{0})+(p_{3}-p_{0})r^{2}t]+
\]
\[
+ik_{4}[(p_{3}-p_{0})r^{2}t-(p_{3}+p_{0})]\}e^{-(\alpha +\beta
)k_{4}^{2}-2i\alpha p_{0}k_{4}}\}\]  \label{90}

After integrating in $\xi $ and $\xi ^{\prime }$, and introducing
the elliptic theta functions representation \cite{Bateman}
\begin{equation}
\theta _{3}\left( u/\tau \right) =\sum\limits_{n=-\infty }^{\infty
}\exp [i\pi (\tau n^{2}+2nu)]  \label{91}
\end{equation}
we find
\begin{equation}
a^{\prime }=a^{(1)}+a^{(2)}  \label{92}
\end{equation}
where
\[
a^{(1)}=\delta ^{4}(p-p^{\prime })\frac{g^{2}\pi ^{2}\sqrt{\pi }\widehat{T}}{%
\widehat{p}_{\Vert }^{2}}\times
\]
\[
\times \int\limits_{0}^{\infty }d\alpha \int\limits_{0}^{\infty
}d\beta
\sinh ^{-1}(\alpha +\beta )\frac{\exp (\frac{\alpha ^{2}}{\alpha +\beta })%
\widehat{p}_{3}^{2}}{\sqrt{\alpha +\beta }}\exp -[\widehat{p}_{\Vert }^{2}+%
\widehat{m}_{e}^{2}]\alpha \exp -\widehat{M}_{W}^{2}\beta \exp -\frac{%
\widehat{p}_{\bot }^{2}}{\coth \alpha +\coth \beta }
\]
\[
\times \left[ \frac{\alpha \widehat{p}_{3}}{\alpha +\beta }[r^{-1}t^{-1/2}(%
\widehat{p}_{0}+\widehat{p}_{3})+rt^{1/2}(\widehat{p}_{3}-\widehat{p}_{0})]+(%
\widehat{p}_{0}^{2}-\widehat{p}_{3}^{2})(r^{-1}t^{-1/2}+rt^{1/2})\right]
\]
\begin{equation}
\times \theta _{3}\left( -2\widehat{p}_{0}\widehat{T}\alpha \left/
i4\pi \widehat{T}^{2}(\alpha +\beta )\right. \right)  \label{93}
\end{equation}
and
\[
a^{(2)}=\delta ^{4}(p-p^{\prime })\frac{g^{2}\pi ^{2}\sqrt{\pi }\widehat{T}}{%
\widehat{p}_{\Vert }^{2}}\times
\]
\[
\times \int\limits_{0}^{\lambda }d\gamma \int\limits_{0}^{\infty
}d\lambda
\sinh ^{-1}(\lambda )\frac{\exp (\frac{\gamma ^{2}}{\lambda })\widehat{p}%
_{3}^{2}}{\sqrt{\lambda }}\exp -[\widehat{p}_{\Vert }^{2}+\widehat{m}%
_{e}^{2}]\gamma \exp -\widehat{M}_{W}^{2}(\lambda -\gamma )\exp -\frac{%
\widehat{p}_{\bot }^{2}}{\coth \gamma +\coth (\lambda -\gamma )}
\]
\begin{equation}
\times [\overline{r}\overline{t}^{1/2}(\widehat{p}_{3}-\widehat{p}_{0})-%
\overline{r}^{-1}\overline{t}^{-1/2}(\widehat{p}_{0}+\widehat{p}_{3})]\frac{%
\partial _{\gamma }\theta _{3}\left( -2\widehat{p}_{0}\widehat{T}\gamma
\left/ i4\pi \widehat{T}^{2}\lambda \right. \right)
}{-2\widehat{p}_{0}} \label{94}
\end{equation}

In Eq. (\ref{94}) we introduced the notation
\begin{equation}
\overline{r}=\exp -2(\lambda -\gamma ),\text{ }\quad
\overline{t}=\exp -2\gamma  \label{95}
\end{equation}

Integrating (\ref{94}) by parts and adding the result with
(\ref{93}), we obtain
\[ \hspace*{-4mm}
a^{\prime }=\delta ^{4}(p-p^{\prime })\frac{g^{2}\pi ^{2}\sqrt{\pi }\widehat{T}}{%
\widehat{p}_{0}\widehat{p}_{\Vert }^{2}}\int\limits_{0}^{\infty
}d\lambda
\frac{\sinh ^{-1}(\lambda )}{\lambda }[\widehat{p}_{3}\sinh (\lambda )+%
\widehat{p}_{0}\cosh (\lambda )]\left[ \exp -[-\widehat{p}_{0}^{2}+\widehat{m%
}_{e}^{2}]\lambda \right] \theta _{3}\left( -2\widehat{p}_{0}\widehat{T}%
\lambda \left/ i4\pi \widehat{T}^{2}\lambda \right. \right)
\]
\[
-\frac{g^{2}\pi ^{2}\sqrt{\pi
}\widehat{T}}{\widehat{p}_{0}\widehat{p}_{\Vert
}^{2}}\int\limits_{0}^{\infty }d\lambda \frac{\sinh ^{-1}(\lambda
)}{\lambda }[\widehat{p}_{3}\sinh (2\lambda )+\widehat{p}_{0}\cosh
(2\lambda )]\left[
\exp -\widehat{M}_{W}^{2}\lambda \right] \theta _{3}\left( 0/i4\pi \widehat{T%
}^{2}\lambda \right)
\]
\[
+\frac{g^{2}\pi ^{2}\sqrt{\pi
}\widehat{T}}{\widehat{p}_{0}\widehat{p}_{\Vert
}^{2}}\int\limits_{0}^{\infty }d\alpha \int\limits_{0}^{\infty }d\beta \frac{%
\sinh ^{-1}(\alpha +\beta )}{\sqrt{\alpha +\beta }}\exp -\widehat{M}%
_{W}^{2}\beta \exp -[\widehat{p}_{\Vert
}^{2}+\widehat{m}_{e}^{2}]\alpha \exp (\frac{\alpha ^{2}}{\alpha
+\beta })\widehat{p}_{3}^{2}
\]
\[
\times \exp -\frac{\widehat{p}_{\bot }^{2}}{\coth \alpha +\coth \beta }%
\left\{ \frac{2\alpha \widehat{p}_{3}\,\widehat{p}_{0}}{\alpha +\beta }%
\left[ \widehat{p}_{3}\cosh (\alpha +2\beta )+\widehat{p}_{0}\sinh
(\alpha +2\beta )\right] -\widehat{p}_{\Vert }^{2}\cosh (\alpha
+2\beta )\right.
\]
\[ \hspace*{-6mm}
+\left[ (\widehat{p}_{\Vert }^{2}+\widehat{m}_{e}^{2}-\widehat{M}_{W}^{2})-%
\frac{2\alpha \widehat{p}_{3}^{2}}{\alpha +\beta }+(\frac{1}{\sinh
^{2}\alpha }-\frac{1}{\sinh ^{2}\beta })\frac{\widehat{p}_{\bot }^{2}}{%
\left[ \coth \alpha +\coth \beta \right] ^{2}}\right] \left[ \widehat{p}%
_{3}\sinh (\alpha +2\beta )+\widehat{p}_{0}\cosh (\alpha +2\beta
)\right]
\]
\begin{equation}
\left. +\left[ \widehat{p}_{3}\cosh (\alpha +2\beta
)+\widehat{p}_{0}\sinh
(\alpha +2\beta )\right] \right\} \theta _{3}\left( -2\widehat{p}_{0}%
\widehat{T}\alpha /\left/ i4\pi \widehat{T}^{2}(\alpha +\beta
)\right. \right)  \label{96}
\end{equation}

To isolate the temperature part of $a^{\prime }$, we have to
subtract the zero-temperature (vacuum) contribution from Eq.
(\ref{96}). With this aim, we use the Jacobi imaginary
transformation \cite{Bateman}
\begin{equation}
\theta _{3}\left( u/\tau \right) =(\sqrt{-i\tau })^{-1}\exp [-i\pi
u^{2}/\tau ]\theta _{3}\left( \frac{u}{\tau }\left/ -\frac{1}{\tau
}\right. \right) ,  \label{97}
\end{equation}
to write
\[
\theta _{3}\left( -2\widehat{p}_{0}\widehat{T}\alpha \left/ i4\pi \widehat{T}%
^{2}(\alpha +\beta )\right. \right) =\frac{\exp
-[\widehat{p}_{0}^{2}\alpha
^{2}/(\alpha +\beta )]}{2\widehat{T}\sqrt{\pi }\sqrt{(\alpha +\beta )}}%
\times
\]
\begin{equation}
\times \left[ 1+\sum\limits_{n=-\infty }^{\infty }\!\text{\/\negthinspace }%
^{^{\prime }}\exp -\left( \frac{n^{2}}{4\widehat{T}^{2}(\alpha +\beta )}+%
\frac{\widehat{p}_{0}\alpha n}{\widehat{T}(\alpha +\beta )}\right)
\right] \label{98}
\end{equation}
\noindent where the symbol $\sum\limits_{n=-\infty }^{\infty }\!$%
\/\negthinspace $^{^{\prime }}$ means that the term $n=0$ was
taken out.

Then, to subtract the vacuum term from $a^{\prime }$ is equivalent
to make in Eq. (\ref{96}) the following substitution
\begin{equation}
\theta _{3}\left( -2\widehat{p}_{0}\widehat{T}\alpha \left/ i4\pi \widehat{T}%
^{2}(\alpha +\beta )\right. \right) \rightarrow \theta _{3}\left( -2\widehat{%
p}_{0}\widehat{T}\alpha \left/ i4\pi \widehat{T}^{2}(\alpha +\beta
)\right.
\right) -\frac{\exp -[\widehat{p}_{0}^{2}\alpha ^{2}/(\alpha +\beta )]}{2%
\widehat{T}\sqrt{\pi }\sqrt{(\alpha +\beta )}},  \label{99}
\end{equation}
\begin{equation}
\theta _{3}\left( -2\widehat{p}_{0}\widehat{T}\lambda \left/ i4\pi \widehat{T%
}^{2}\lambda \right. \right) \rightarrow \theta _{3}\left( -2\widehat{p}_{0}%
\widehat{T}\lambda \left/ i4\pi \widehat{T}^{2}\lambda \right. \right) -%
\frac{\exp -[\widehat{p}_{0}^{2}\lambda ]}{2\widehat{T}\sqrt{\pi }\sqrt{%
\lambda }},  \label{100}
\end{equation}
\begin{equation}
\theta _{3}\left( 0/i4\pi \widehat{T}^{2}\lambda \right)
\rightarrow \theta
_{3}\left( 0/i4\pi \widehat{T}^{2}\lambda \right) -\frac{1}{2\widehat{T}%
\sqrt{\pi }\sqrt{\lambda }}.  \label{101}
\end{equation}

\begin{figure}
\includegraphics{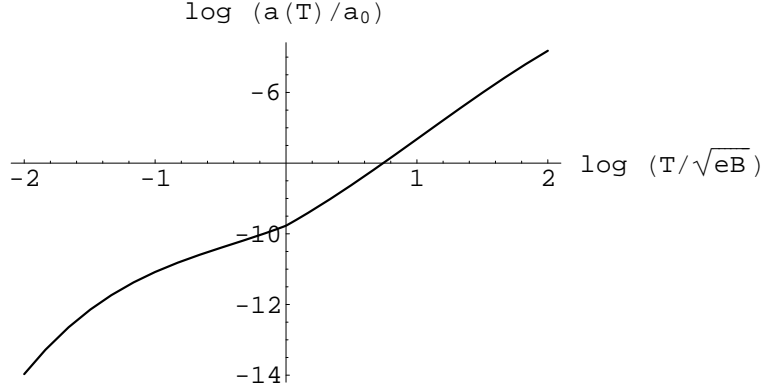}
\caption{\label{a-coefT}Plot of $\log (a(T)/a_{0})$ vs. $\log
(\widehat{T})$ for field range $10^{-2}T\leq \sqrt{eB}\leq
10^{2}T$ and parameter values
$|\widehat{p}_{\parallel}|=|\widehat{p}_{\perp }|=10^{-1} \sqrt{2}
\widehat{m}_{e}$, $\widehat{M}_{W}=10^{5} \widehat{m}_{e}$,
$\widehat{m}_{e}=0.1$}
\end{figure}

which leads to
\begin{eqnarray}
a^{\prime }(T) &=&\delta ^{4}(p-p^{\prime })\frac{g^{2}\pi ^{2}\sqrt{\pi }%
\widehat{T}}{\widehat{p}_{0}\widehat{p}_{\Vert
}^{2}}\{\int\limits_{0}^{\infty }d\lambda \frac{\sinh
^{-1}(\lambda )}{\lambda }[\widehat{p}_{3}\sinh
(\lambda )+\widehat{p}_{0}\cosh (\lambda )]\left[ \exp -[-\widehat{p}%
_{0}^{2}+\widehat{m}_{e}^{2}]\lambda \right]   \nonumber \\
&&\left[ \theta _{3}\left( -2\widehat{p}_{0}\widehat{T}\lambda /
i4\pi
\widehat{T}^{2}\lambda \right) -\frac{\exp -[\widehat{p}_{0}^{2}\lambda ]}{2%
\widehat{T}\sqrt{\pi }\sqrt{\lambda }}\right]   \nonumber \\
&&-\int\limits_{0}^{\infty }d\lambda \frac{\sinh ^{-1}(\lambda )}{\lambda }[%
\widehat{p}_{3}\sinh (2\lambda )+\widehat{p}_{0}\cosh (2\lambda )]
\nonumber
\\
&&\times \left[ \exp -\widehat{M}_{W}^{2}\lambda \right] \left[
\theta
_{3}\left( 0/i4\pi \widehat{T}^{2}\lambda \right) -\frac{1}{2\widehat{T}%
\sqrt{\pi \lambda }}\right]   \nonumber \\
&&+\int\limits_{0}^{\infty }d\alpha \int\limits_{0}^{\infty }d\beta \frac{%
\sinh ^{-1}(\alpha +\beta )}{\sqrt{\alpha +\beta }}\exp -[\widehat{M}%
_{W}^{2}\beta ]\exp -[\widehat{p}_{\Vert
}^{2}+\widehat{m}_{e}^{2}]\alpha
\nonumber \\
&&\times \exp (\frac{\alpha ^{2}}{\alpha +\beta })\widehat{\,p}_{3}^{2}\exp -%
\frac{\widehat{p}_{\bot }^{2}}{\coth \alpha +\coth \beta }  \nonumber \\
&&\times (\left[ \frac{2\alpha
\widehat{p}_{3}\,\widehat{p}_{0}}{\alpha
+\beta }+1\right] \left[ \widehat{p}_{3}\cosh (\alpha +2\beta )+\widehat{p}%
_{0}\sinh (\alpha +2\beta )\right] -\widehat{p}_{\Vert }^{2}\cosh
(\alpha
+2\beta )  \nonumber \\
&&+\left[ (\widehat{p}_{\Vert }^{2}+\widehat{m}_{e}^{2}-\widehat{M}_{W}^{2})-%
\frac{2\alpha \widehat{p}_{3}^{2}}{\alpha +\beta }+(\frac{1}{\sinh
^{2}\alpha }-\frac{1}{\sinh ^{2}\beta })\frac{\widehat{p}_{\bot }^{2}}{%
\left[ \coth \alpha +\coth \beta \right] ^{2}}\right]   \nonumber \\
&&\times \left[ \widehat{p}_{0}\sinh (x+2y)+\widehat{p}_{3}\cosh
(x+2y)\right] )  \nonumber \\
&&\times \left[ \theta _{3}\left(
-2\widehat{p}_{0}\widehat{T}\alpha / i4\pi
\widehat{T}^{2}(\alpha +\beta )\right) -\frac{\exp -[\widehat{p}%
_{0}^{2}\alpha ^{2}/(\alpha +\beta )]}{2\widehat{T}\sqrt{\pi (\alpha +\beta )%
}}\right]\}   \label{102}
\end{eqnarray}

Eq. (\ref{102}) gives the thermal part of the $a^{\prime }$ taking
into account all Landau levels. To compare the thermal and the LLL
contributions we plot $\log a(T)/a_{0}$ vs. $\log
\widehat{T}^{-1}$ in Fig. 3.

$ $From Fig. 3 we see that in all the range of temperatures
considered, the thermal contribution is consistently smaller in
several orders than the LLL contribution. This result, along with
those obtained in Figs. 1 and 2, imply that $\varkappa \approx 1$,
thereby validating the LLL approximation in the strong and
weakly-strong field cases ($M_{W}\gg \sqrt{B}\gg m_{e},\left|
\mathbf{p}\right| $, $eB\gtrsim $ $T^{2}$). In addition, we would
like to underline the consistency of these numerical results with
the arguments used in Sec.~\ref{strong-b-selfen} (see discussion
after Eq. (\ref{44})), in the sense that the finite temperature
part should be much smaller than the vacuum contribution.

\section{Applications to Cosmology and Astrophysics}\label{applicat}

Two natural environments where the results of the present paper
can be of interest are, on one side, stars possessing large
magnetic fields (neutron stars, magnetars, etc), and, on the
other, the early universe, which presumably was permeated by large
primordial magnetic fields that eventually became the seeds
\cite{Grasso}, \cite{enq} of the galactic fields that are observed
today at scales of $100$ Kpc  \cite{Kronberg} and larger.

The effects of primordial magnetic fields on neutrino propagation
are of special interest during the neutrino decoupling era, on
which a large number of these particles escaped the electroweak
plasma with the possibility to develop significant oscillations.
The strength of the primordial magnetic field in the neutrino
decoupling era can be estimated from the constraints derived from
the successful nucleosynthesis prediction of primordial $^{4}He$
abundance \cite{He}$,$ as well as on the neutrino mass and
oscillation limits \cite{Enqvist}. These constraints, together
with the energy equipartition principle, lead to the relations
\begin{equation}
m_{e}^{2}\leq eB\leq m_{\mu }^{2},\qquad B/T^{2}\sim 2,
\label{105}
\end{equation}
with $m_{\mu }$ being the muon mass. Then, it is reasonable to
assume that between the QCD phase transition epoch and the end of
nucleosynthesis a primordial magnetic field in the above range
could have been present \cite {Elizalde}.

Unlike the stellar material, whose density can be of order one, we
know that the early universe was almost charge symmetric, with a
particle-antiparticle asymmetry of only $10^{-9}-10^{-10}$.
Therefore, when considering possible consequences of our results
for the early universe, we should restrict the discussion to the
neutral case ($\mu =0$).

For the neutral case, it is known that at weak field all the
corrections in the neutrino energy density, whether they depend on
the field and/or temperature, are second order in the Fermi
coupling constant \cite{Raffelt}, \cite{McKeon}, \cite{Weak-F},
therefore negligible small. Nevertheless, it can be seen from Eqs.
(\ref{63})-(\ref{64}) that if sufficiently strong magnetic fields
were present in the primeval plasma, they would yield corrections
to the energy density that are linear in the Fermi coupling
constant. Moreover, as shown in Sec.~\ref{numerical}, the
strong-field approximation that led to these corrections is
reliable even for magnetic fields in a more realistic range (i.e.
as those satisfying condition (\ref{105})).

Notice that a field satisfying (\ref{105}) would be effectively
strong with respect to the electron-neutrino, but weak for the
remaining neutrino flavors. If such a field existed during the
decoupling era, it could significantly affect the $\nu
_{e}\leftrightarrow \nu _{\mu },\nu _{\tau }$
and $\nu _{e}\leftrightarrow \nu _{s}$ resonant oscillations \cite{Australia}%
, as the field would differently modify the energy of $\nu _{e}$
compared to those of $\nu _{\mu }$, $\nu _{\tau }$ and $\nu _{s}.$
The interesting new thing here is that despite these oscillations
would take place in an essentially neutral medium, because of the
strong field they will be as significant as those produced by the
MSW mechanism \cite{msw} in a dense medium. A peculiarity of the
strong-field effects on the neutrino energy density is to give
rise to anisotropic resonant oscillations. That is, the
oscillation probability depends on the direction of the neutrino
propagation with respect to the magnetic field.

Another interesting question related to primordial magnetic fields
is whether they influenced neutrino propagation prior to the
electroweak phase transition, since some of the mechanisms of
primordial magnetic field generation allow their existence at very
early epochs \cite {Elect-W}, \cite{Grasso}. Before the
electroweak transition a primordial magnetic field could only
exist in the form of a U(1) hypermagnetic field \cite{Laine}. Any
non-Abelian ``magnetic'' field would decay at high temperatures
because it would acquire a non-perturbative infrared magnetic mass
~$g^{2}T$. The implications of primordial hypermagnetic fields in
neutrino propagation before the electroweak phase transition have
been studied in Refs.~\cite{Cannellos}, \cite{Ayala}. In
\cite{Cannellos} the strong-field effects led to a large
anisotropy in the leptons' energies. The anisotropy was due to the
degeneracy in the energy, which in the leading order does not
depend on the transverse momentum component. It is in contrast
with the magnetic field case, which takes place after the
electroweak phase transition, where, although an anisotropy in the
neutrino energy is also present, no degeneracy occurs (see
Eqs.~\ref{63})-(\ref{64})). This different behavior is a direct
consequence, on one hand, of the nonzero neutrino hypercharge and
hence of the minimally coupling of these particles to the
hypermagnetic field; and on the other, of their electrical
neutrality, which implies that they can interact with a magnetic
field through radiative corrections only.

We underline that the anisotropy in the energy spectrum of
neutrinos in background fields could provide an independent way to
verify the existence of primordial magnetic fields, since a
field-induced anisotropy would be reflected as an imprint in a yet
undetected and elusive cosmic background of neutrinos produced
during the decoupling era.

For astrophysical applications we should turn our attention to the
star interiors characterized by high densities of charged
particles under strong magnetic fields. As shown in
Sec.~\ref{index-ref}, in the charged medium case the strong-field
approximation gives rise to a modification of the kinetic part of
the neutrino energy (first term in Eq.(\ref{68})), that is not
present in leading order at weak field (Eq.~(\ref{71}))
\cite{Nieves}. Due to this new term the largest field-dependent
contribution to the neutrino kinetic energy comes from neutrinos
propagating perpendicularly to the field.

It is worth noticing that the anisotropic term associated with the
effective magnetic moment in the dispersion relations (\ref{68})
and (\ref {71}) changes sign when neutrinos reverse their
direction of motion. On the other hand, the kinetic energy term in
(\ref{68}) remains unchanged when the neutrinos moves in opposite
direction, even though it depends on the direction of propagation
of the neutrinos with respect to the magnetic field. As known
\cite{Kusenko}, the anisotropy associated to the magnetic-moment
contribution to the neutrino energy can be relevant for a possible
explanation of the peculiar high pulsar velocities. It would be
interesting to consider the combined effects of the two different
anisotropies to understand whether they could affect the dynamics
of proto-neutron stars.

Another possible ground of applications of our results is
supernova neutrinos. Core collapse supernovae are dominated by
neutrinos and their transport properties. In addition, the
observation of neutrinos from supernova, which is essential to
confirm the basic picture of supernova explosion, can be affected
by neutrino oscillations. As we have pointed out in this paper,
neutrino transport and oscillations can be both modified by the
presence of a strong magnetic field. Magnetic fields as strong as
$10^{14}$ to $10^{16}$ G could exist in the first seconds of
neutrino emission inside the supernova core \cite{Thompson}. Thus,
the electron-neutrino energy spectrum found in this paper, within
the strong-field limit for the charged medium, should be
considered for any study of neutrino oscillations under those
conditions.

\section{Concluding Remarks}\label{conclusions}

In this paper we carried out a thorough study of the propagation
of neutrinos in strongly magnetized neutral and charged media. We
started from the most general structure of the neutrino
self-energy in a magnetic field, expressing it as the sum of four
independent covariant terms with coefficients that are functions
of the physical variables of the theory and whose values depend on
the approximation considered. General expressions of the four
coefficients at one-loop approximation were given in Eqs.
(\ref{25}-\ref{28}).

The coefficients were then calculated in the strong-field limit
using the LLL aproximation for the electrons. The LLL was assumed
to be valid in the parameter range $M_{W}\gg \sqrt{B}\gg
m_{e},\left| \mathbf{p}\right| $, $eB\gtrsim $ $T^{2}$. To justify
it one should keep in mind that under these conditions most
electrons would not have enough energy to overcome the gap between
the Landau levels. Hence, they will be mainly confined to their
lower levels and the leading contribution would come from the LLL.
This assumption was also corroborated for the above parameters'
range by numerical calculations summing in all Landau levels.

The dispersion relation of the neutrinos was written as a function
of the four coefficients of the self-energy structures, allowing
in this way to straightforwardly obtain the neutrino's energy in
the strong-field limit for each physical case.

In concordance with results previously obtained in charged media
at weak fields \cite{Nieves},%
\cite{Weak-F}, in the strong-field case an energy term associated
with the interaction between the magnetic field and the effective
magnetic moment was also found at leading order in $G_{F}$. This
interaction energy disappears in the neutral medium, since in a
charged-symmetric plasma the contribution to the effective
magnetic moment coming from electrons and positrons cancels out.

A main outcome of our investigation was to show that in strongly
magnetized systems a term of different nature emerges in both
charged and neutral media. The new term, which is linear in the
magnetic field and of first order in $G_{F}$, enters as a
correction to the neutrino kinetic energy in the presence of a
strong-magnetic field. This correction is present even in a
strongly magnetized vacuum, since it is related to the vacuum part
($T=0,$ $\mu =0$) of the neutrino self-energy at $B\neq 0$.

A characteristic of the field-dependent corrections to the
neutrino energy is that they produce an anisotropic index of
refraction, since neutrinos moving along different directions have
different field-dependent dispersion relations. We should
underline that while the magnetic moment interaction term produces
a maximum field effect for neutrinos propagating along the field
lines, the field correction to the kinetic energy does not
contribute to those propagation modes, but on the contrary, the
maximum kinetic-energy effect takes place for neutrinos
propagating perpendicularly to the field direction. We stress that
the anisotropy does not differentiate between neutrinos and
antineutrinos.

The charged medium results reported in the current work can be of
interest for the astrophysics of neutrinos in stars with large
magnetic fields. On the other hand, our finding for the neutral
medium can have applications in cosmology, if the existence of
high primordial magnetic fields is finally confirmed. Contrary to
some authors' belief \cite{Weak-F},\cite{Dolgov} that, regardless
of the field intensity, the neutrino dispersion relation in the
early universe is well approximated by the dispersion relation in
the zero field medium, our results indicate that strong, and even
weakly strong, magnetic fields can give rise to a contribution to
the neutrino energy that is several orders larger than the pure
thermal contribution.

The field-dependent correction to the neutrino energy in a neutral
medium with strong magnetic field can have an impact in neutrino
flavor-oscillations in the primeval plasma \cite{Australia} and
therefore affect primordial nucleosynthesis. Hence, this new
effect could be important to establish possible limits to the
strength of the primordial magnetic field.

\medskip

\textbf{Acknowledgments}

EE is indebted to the members of the Mathematics Dept. of MIT, and
specially to Dan Freedman, for the warm hospitality. EJF and VI
would like to thank the Institute of Space Studies of Catalonia
(Spain), where part of this work was accomplished, for the warm
hospitality during their visit. The work of EE has been supported
by DGICYT (Spain), contract PR-2003-0352 and project
BFM2003-00620. The work of EJF and VI was supported by NSF-grant
PHY-0070986.

\end{document}